\documentclass[fleqn,usenatbib]{mnras}

\usepackage[T1]{fontenc}

\usepackage{graphicx}	
\usepackage{amsmath}	
\usepackage{amssymb}	
\usepackage[normalem]{ulem}
\usepackage{bm}
\usepackage{pdflscape}

\usepackage{newtxtext,newtxmath}

\title[Forecasting magnification for the LSST]{Forecasting the potential of weak lensing magnification to enhance LSST large-scale structure analyses}

\author[C. Mahony et al.]{
Constance Mahony$^{1, 2}$\thanks{E-mail: mahony@astro.rub.de},
Maria Cristina Fortuna$^{3}$,
Benjamin Joachimi$^{1}$,
Andreas Korn$^{1}$,
\newauthor
Henk Hoekstra$^{3}$,
Samuel J. Schmidt$^{4}$
and the LSST Dark Energy Science Collaboration
\\
$^{1}$Department of Physics and Astronomy, University College London, Gower Street, London WC1E 6BT, UK\\
$^{2}$Ruhr University Bochum, Faculty of Physics and Astronomy, Astronomical Institute (AIRUB), German Centre for Cosmological Lensing, \\ 44780 Bochum, Germany\\
$^{3}$Leiden Observatory, Leiden University, PO Box 9513, Leiden, NL-2300 RA, The Netherlands\\ 
$^{4}$Department of Physics, University of California, One Shields Ave., Davis, CA 95616, USA}

\date{Accepted XXX. Received YYY; in original form ZZZ}

\pubyear{2021}

\begin{document}
\label{firstpage}
\pagerange{\pageref{firstpage}--\pageref{lastpage}}
\maketitle

\begin{abstract}
  Recent works have shown that weak lensing magnification must be included in upcoming large-scale structure analyses, such as for the Vera C. Rubin Observatory Legacy Survey of Space and Time (LSST), to avoid biasing the cosmological results. In this work we investigate whether including magnification has a positive impact on the precision of the cosmological constraints, as well as being necessary to avoid bias. We forecast this using an LSST mock catalog and a halo model to calculate the galaxy power spectra. We find that including magnification has little effect on the precision of the cosmological parameter constraints for an LSST galaxy clustering analysis, where the halo model parameters are additionally constrained by the galaxy luminosity function. In particular, we find that for the LSST gold sample ($i < 25.3$) including weak lensing magnification only improves the galaxy clustering constraint on $\Omega_{\rm{m}}$ by a factor of 1.03, and when using a very deep LSST mock sample ($i<26.5$) by a factor of 1.3. Since magnification predominantly contributes to the clustering measurement and provides similar information to that of cosmic shear, this improvement would be reduced for a combined galaxy clustering and shear analysis. We also confirm that not modelling weak lensing magnification will catastrophically bias the cosmological results from LSST. Magnification must therefore be included in LSST large-scale structure analyses even though it does not significantly enhance the precision of the cosmological constraints. 
\end{abstract}




\section{Introduction}

As light from distant galaxies travels towards telescopes it is deflected gravitationally by intervening matter. This means that galaxy images appear distorted. On average, the distortions to individual galaxy images are very small, but when combined, they can be used to statistically map the matter distribution in the universe. This technique is called weak gravitational lensing. 

Weak gravitational lensing distorts both the shape and size of galaxy images. Statistical measurements of the shape distortions are referred to as cosmic shear, and statistical measurements of the size distortions are referred to as magnification. Making a magnification measurement of the matter distribution in the Universe, which directly uses size information is challenging because there is a large intrinsic variation in the sizes of galaxies, and it is more prone to serious systematics \citep{hoekstra2017}. However, \cite{Schmidt_2011} achieved a simplified magnification measurement using the joint distribution of galaxy sizes and magnitudes, and there are developing techniques which anchor the size distribution using the fundamental plane of galaxies \citep{Huff_2013,Freudenburg:2019dyh}. Most magnification analyses therefore focus on making a magnification measurement using galaxy number density information \citep{Scranton:2005ci,Myers:2005ub,2009Hildebrandt}. In a flux limited survey, distortions to the sizes of galaxy images affect the observed number density of galaxies for two reasons:
\begin{enumerate}
  \item Since surface brightness is conserved by lensing if the observed size of a galaxy is increased, so is its observed flux. This means that galaxies previously too faint to be observed by a galaxy survey become observable. The number density of galaxies is increased.
  \item It is not only the observed size of individual galaxies that is increased by magnification, but the observed size of the whole patch of sky behind the lens. This means that the observable separation between galaxies behind the lens increases and there is a dilution in the number density of galaxies.
\end{enumerate}
These two effects compete and contribute to an overall fluctuation in the number density of galaxies, as a result of weak lensing magnification (for the associated equations see section \ref{sec:2D power spectra}). Here we are concerned with how magnification can probe the total matter distribution, but it can also be used to constrain the mass of galaxy clusters (e.g. \citealt{Tudorica:2017crp}).

Weak lensing using cosmic shear has been a highly successful technique. In recent years there have been increasingly precise results using cosmic shear from galaxy surveys such as the Kilo Degree Survey (KiDS) \citep{vanUitert:2017ieu,Joudaki:2017zdt,Hildebrandt:2018yau,asgariKiDS:2020suj}, the Dark Energy Survey (DES) \citep{Abbott:2017wau,Troxel:2017xyo,AmonDES:2021bvc,seccoDES:2021vln} and the Hyper Suprime-Cam Survey (HSC) \citep{Hikage:2018qbn}. Weak lensing magnification has not been included in standard weak lensing analyses to date. All that has been included is the sensitivity of results to including a simplified magnification model \citep{Abbott:2018xao}. The reasoning is that magnification provides similar information to that of cosmic shear and has a poorer signal-to-noise ratio \citep{Bartelmann:2010fz}. However, due to improvements in statistical precision, recent works have shown that cosmological results from upcoming surveys such as the Vera C. Rubin Observatory Legacy Survey of Space and Time (LSST) and \textit{Euclid} will be biased if the effects of weak lensing magnification are not included \citep{Duncan:2013haa, Cardona:2016qxn, Lorenz:2017iez, Thiele:2019fcu}. 

These works have shown that magnification must be included in future surveys to avoid bias, but the aim of this work is to determine whether including magnification as a complementary probe can also improve the final precision of the LSST weak lensing results. \cite{Duncan:2013haa} and \cite{Lorenz:2017iez} found no increase in precision from including magnification in a weak lensing analysis, however LSST is a special case, because it is a very deep ground based galaxy survey. This means that there will be a lot of very faint, small and distant galaxies, which will be poorly resolved. It will therefore not be possible to measure the shape of these galaxies, but it may be possible to count them for a weak lensing magnification analysis. This means that the potentially usable sample size for weak lensing magnification is significantly larger than that for cosmic shear, and as such it is worth investigating magnification's potential as a complementary probe in the case of LSST \footnote{\cite{Lorenz:2017iez} also considered LSST specifically, but did not include systematics or explore departures from the gold sample used for cosmic shear.}. Particularly, as \cite{Nicola:2019yiw} showed that even with only approximately 100 square degrees, deep samples are already sensitive to magnification.

In summary, we wish to determine the effect of including weak lensing magnification on the precision of the final constraints from LSST weak lensing. We determine this using the Fisher matrix formalism introduced in section \ref{sec:fisher_analysis}. We then describe the modelling of the observables (weak lensing power spectra and the galaxy luminosity function) in sections \ref{sec:weak_lensing_observables} and \ref{sec:galaxy LF}. We describe the details of our LSST specific survey modelling in section \ref{sec:survey modelling}; and present our results and conclusions in sections \ref{sec:results} and \ref{sec:Conclusions}. We verify the stability of our Fisher matrices in appendix \ref{sec: Fisher matrix stability}.

\section{Fisher Analysis} \label{sec:fisher_analysis}
The Fisher Information matrix summarises the expected curvature of the log-Likelihood function around its maximum,
\begin{equation}
  F_{ij} = \bigg \langle \frac{- \partial^2 \ln L}{\partial \theta_i \partial \theta_j} \bigg \rangle \ ,
\end{equation}
where $L$ is the likelihood and $\theta_i$ is a model parameter. If the likelihood function is sharply peaked for a given parameter, the parameter is tightly constrained by the data \citep{Dodelson}. The marginal uncertainty on the model parameter $\theta_i$ can be calculated from the Fisher matrix as:
\begin{equation}
  \Delta \theta_i \geq \sqrt{(F^{-1})_{ii}} \ .
\end{equation}
The greater than or equal relation is in reference to the Cram\'er-Rao inequality, which specifies that the Fisher matrix gives the minimum possible uncertainty on an unbiased model parameter \citep{Tegmark:1996bz}. 

The Fisher information matrix can be calculated without data and is therefore a useful tool for forecasting best case parameter constraints. In the case of a Gaussian likelihood function and a parameter independent covariance matrix the Fisher matrix is given by:
\begin{equation} \label{eq:fisher_sum}
  F_{ij} = \sum_{\ell} \frac{\partial C_\ell}{\partial \theta_i} \ \mathrm{Cov}^{-1} \ \frac{\partial C_\ell}{\partial \theta_j} \ ,
\end{equation}
where $C$ is the theory datavector and $\mathrm{Cov}$ is the associated covariance \citep{Tegmark:1996bz}. In this work we consider two component Fisher matrices, which we then add together since they concern separate observables: the Fisher matrix where the theory datavector consists of galaxy clustering or galaxy clustering and cosmic shear (detailed in section \ref{sec:weak_lensing_observables}), and the Fisher matrix where the theory datavector consists of the galaxy luminosity function (detailed in section \ref{sec:galaxy LF}). There may be a small correlation between the observables due to cosmic variance, but we do not consider this in this forecast. The associated covariances are detailed in sections \ref{sec:2pt cov} and \ref{sec:CLF fiducial values} respectively.

\section{Weak Lensing Observables} \label{sec:weak_lensing_observables}

The two observable quantities used in this weak lensing analysis are the shape, often referred to as ellipticity, and the number density of galaxy images. Since weak lensing is a local effect, the mean ellipticity $\epsilon$ and fluctuation in the number density of galaxies $n$, resulting from weak lensing, is equal to zero when averaged over large scales in the absence of systematics. Therefore, the key statistical quantity used in weak lensing analyses is the two-point correlation function. There are three two-point correlation functions commonly considered in large-scale structure and weak lensing analyses: cosmic shear (ellipticity-ellipticity), angular galaxy clustering (number density-number density) and galaxy-galaxy lensing (number density-ellipticity). In this work we focus on angular galaxy clustering as an individual probe (section \ref{sec:clustering results} and \ref{sec:bias results}), and also consider a combined clustering and shear analysis, where the analyses occur on separate patches of sky (section \ref{sec:shear calibration}).

The Fourier space two-point correlation function for angular galaxy clustering is given by:
\begin{equation}\label{eq:n-n}
  \langle \tilde{n}^i(\boldsymbol{\ell}) \tilde{n}^j(\boldsymbol{\ell^\prime}) \rangle = (2\pi)^2 \delta^{(2)}(\boldsymbol{\ell}+\boldsymbol{\ell^\prime})C_{\mathrm{nn}}^{ij}(\ell) \ ,
\end{equation}
where $\tilde{n}$ is the Fourier transform of the number density contrast, $\boldsymbol{\ell}$ is the angular frequency, $\delta^{(2)}$ is the two-dimensional Dirac delta function and $C_{\mathrm{nn}}^{ij}$ is the projected number density power spectrum between redshift bins $i$ and $j$ \citep{Joachimi:2010}. It is useful to work in Fourier space because it simplifies linking to the theory predictions. The galaxy samples used for weak lensing are often split into redshift bins; a technique called redshift tomography. This binning enables weak lensing to probe the evolution of the power spectrum with time, through auto- and cross-correlations between the different redshift bins, and hence study the expansion of the universe and dark energy.

The Fourier transform of the two-point correlation function for cosmic shear is given by:
\begin{equation} \label{eq:e-e}
  \langle \tilde{\epsilon}^i(\boldsymbol{\ell}) \tilde{\epsilon}^j(\boldsymbol{\ell^\prime}) \rangle = (2\pi)^2 \delta^{(2)}(\boldsymbol{\ell}+\boldsymbol{\ell^\prime})C_{\mathrm{\epsilon\epsilon}}^{ij}(\ell) \ ,
\end{equation}
where $\tilde{\epsilon}$ is the Fourier transform of the ellipticity and $C_{\mathrm{\epsilon\epsilon}}^{ij}$ is the projected ellipticity power spectrum between redshift bins $i$ and $j$ .

\subsection{2D power spectra} \label{sec:2D power spectra}

The key quantities in equations \ref{eq:n-n} and \ref{eq:e-e}, are the two-dimensional (2D) power spectra $C_{\mathrm{nn}}$ and $C_{\mathrm{\epsilon\epsilon}}$. These are the observables we model and include in our Fisher matrix theory datavector, see section \ref{sec:fisher_analysis}. 

In this work we model the 2D observable power spectra $C_{\mathrm{nn}}$ and $C_{\mathrm{\epsilon\epsilon}}$ by breaking them down into their constituent parts. The observed ellipticity of a galaxy comes from a combination of the intrinsic ellipticity of the galaxy before it is lensed $\epsilon_I$ (the intrinsic alignment, see \citealt{Joachimi:2015mma, Troxel:2014dba} for reviews), the distortion of the shape by weak lensing shear $\gamma_G$, and a random uncorrelated component $\epsilon_{\rm{rnd}}$ which accounts for the randomness in the intrinsic ellipicity of galaxies,
\begin{equation} \label{eq:epsilon_contirbutions}
  \epsilon^i(\bm{\theta}) = \gamma^i_G(\bm{\theta})+\epsilon^i_I(\bm{\theta})+\epsilon^i_{\rm{rnd}}(\bm{\theta}) \ ,
\end{equation}
where $i$ denotes the redshift bin. The observed number density of galaxies comes from a combination of the number density fluctuation of galaxies as a result of galaxy clustering $n_g$, the distortion to the number density from weak lensing magnification $n_m$, and a random component $n_{\rm{rnd}}$ which accounts for the shot noise contribution,
\begin{equation} \label{eq:n_constributions}
  n^i(\bm{\theta}) = n^i_g(\bm{\theta})+n^i_m(\bm{\theta})+n^i_{\rm{rnd}}(\bm{\theta}) \ .
\end{equation}

In terms of the Fourier space 2D power spectra the uncorrelated random components lead to noise power spectra, and separating out the remaining contributions gives:
\begin{equation}
  \begin{split}
  C_{\mathrm{\epsilon\epsilon}}^{ij}(\ell) = C_{\mathrm{GG}}^{ij}(\ell) + C_{\mathrm{IG}}^{ij}(\ell) + C_{\mathrm{IG}}^{ji}(\ell) + C_{\mathrm{II}}^{ij}(\ell) \ , \\\
  C_{\mathrm{nn}}^{ij}(\ell) = C_{\mathrm{gg}}^{ij}(\ell) + C_{\mathrm{gm}}^{ij}(\ell) + C_{\mathrm{gm}}^{ji}(\ell) + C_{\mathrm{mm}}^{ij}(\ell) \ , \\
  \end{split} \label{eq: observables}
  \end{equation}
where G represents ellipticity from weak lensing shear, I ellipticity from the intrinsic alignment of galaxies, g number density fluctuations as a results of intrinsic galaxy clustering and m number density fluctuations as a result of weak lensing magnification.

We compute all these two-dimensional power spectra $C_{\mathrm{ab}}$ from their associated three-dimensional power spectra $P_{\mathrm{ab}}$ using the Limber approximation in Fourier space \citep{Kaiser1992ApJ...388..272K}:
\begin{equation} \label{eq:limber}
  \begin{split}
  & C_{\mathrm{GG}}^{ij}(\ell) = \int_0^{\chi_{\rm{hor}}}\mathrm{d}\chi\frac{q^i(\chi)q^j(\chi)}{f_K^2(\chi)}P_{\rm{\delta\delta}}\bigg(k=\frac{\ell+1/2}{f_K(\chi)}, \chi\bigg) \ , \\
  & C_{\mathrm{IG}}^{ij}(\ell) = \int_0^{\chi_{\rm{hor}}}\mathrm{d}\chi\frac{p^i(\chi)q^j(\chi)}{f_K^2(\chi)}P_{\rm{I \delta}}\bigg(k=\frac{\ell+1/2}{f_K(\chi)}, \chi\bigg) \ , \\
  & C_{\mathrm{II}}^{ij}(\ell) = \int_0^{\chi_{\rm{hor}}}\mathrm{d}\chi\frac{p^i(\chi)p^j(\chi)}{f_K^2(\chi)}P_{\rm{II}}\bigg(k=\frac{\ell+1/2}{f_K(\chi)}, \chi\bigg) \ , \\
  & C_{\mathrm{gg}}^{ij}(\ell) = \int_0^{\chi_{\rm{hor}}}\mathrm{d}\chi\frac{p^i(\chi)p^j(\chi)}{f_K^2(\chi)}P_{\rm{gg}}\bigg(k=\frac{\ell+1/2}{f_K(\chi)}, \chi\bigg) \ , \\
  & C_{\mathrm{gm}}^{ij}(\ell) = 2(\alpha^{j}-1)C_{\mathrm{gG}}^{ij}(\ell) \ , \\
  & C_{\mathrm{mm}}^{ij}(\ell) = 4(\alpha^{i}-1)(\alpha^{j}-1)C_{\mathrm{GG}}^{ij}(\ell) \ , \\
  & C_{\mathrm{gG}}^{ij}(\ell) = \int_0^{\chi_{\rm{hor}}}\mathrm{d}\chi\frac{p^i(\chi)q^j(\chi)}{f_K^2(\chi)}P_{\rm{g\delta}}\bigg(k=\frac{\ell+1/2}{f_K(\chi)}, \chi\bigg) \ , \\
 \end{split}
\end{equation}
where $\chi$ is the comoving distance, $f_K(\chi)$ is the comoving angular diameter distance and $p^i(\chi)$ is the probability distribution of galaxies in redshift bin $i$. $q^i(\chi)$ is a weight function given by,
\begin{equation}
  q^i(\chi) = \frac{3 H_0^2\Omega_{\rm{m}}}{2c^2}\frac{f_K(\chi)}{a(\chi)} \int_\chi^{\chi_{\rm{hor}}} \mathrm{d}\chi' p^i(\chi')\frac{f_K(\chi'-\chi)}{f_K(\chi')} \ ,
\end{equation}
where $H_0$ is the Hubble constant, $\Omega_{\mathrm{m}}$ the matter density parameter, and $a(\chi)$ the scale factor (for further details see \citealt{Bartelmann:1999yn}). The calculation of the three-dimensional power spectra $P_{\mathrm{ab}}$ is detailed in the following section.

Equation (\ref{eq:limber}) shows that the 2D power spectra associated with magnification $C_{\mathrm{gm}}$ and $C_{\mathrm{mm}}$ can be computed from the 2D power spectra associated with weak lensing shear $C_{\mathrm{gG}}$ and $C_{\mathrm{GG}}$ using $\alpha^{i}$ the faint end slope of the number counts in redshift bin $i$. We discuss the galaxy luminosity function in section \ref{sec:galaxy LF} but detail the relationship between the magnification and shear power spectra here. 

As mentioned previously, weak lensing magnification contributes to fluctuations in the number density of galaxies $n$. If the number density of galaxies above the flux limit $f$ is $N_0(>f)$, magnification alters the number density of sources as: 
\begin{equation}
N(>f) = \frac{1}{\mu}N_0(>f/\mu) \ ,
\end{equation}
where $N(>f)$ is the observed cumulative number density of sources and $\mu$ is the local magnification factor \citep{Bartelmann:1999yn}. If the cumulative number density of galaxies is assumed to follow a power law $N_0(>f) = kf^{-\alpha}$ near the flux limit of the survey then, 
\begin{equation}
N(>f) = \frac{1}{\mu}k\bigg(\frac{f}{\mu}\bigg)^{-\alpha} = N_0(>f)\mu^{\alpha - 1},
\end{equation}
where $\alpha$ is equivalent to $\alpha^{i}$ mentioned in the previous paragraph. This means the fluctuation in the observed number density of galaxies as a result of magnification $n_m$ is given by,
\begin{equation} \label{eq:alpha}
\begin{split}
n_m &= \frac{N(>f)-N_0(>f)}{N_0(>f)} = \mu^{\alpha-1} - 1 \approx (1+2\kappa)^{\alpha-1} - 1 \\
 &\approx 2(\alpha - 1)\kappa ,
\end{split}
\end{equation}
where the weak lensing limit $\mu \approx 1 + 2\kappa$ has been employed.

\subsection{3D power spectra} \label{sec:3D power spectra}

The fundamental ingredient for the construction of all of the 3D power spectra $P_{\rm{ab}}$ in eq. (\ref{eq:limber}) is the matter power spectrum $P_{\rm{\delta\delta}}$. It summarises the clustering of matter in the universe and can be derived numerically using the Boltzmann equations and the primordial power spectrum predicted by inflation. For the other power spectra, we can only rely on an effective description, which we detail in this section.
 
In this work, we compute $P^\mathrm{lin}_{\rm{\delta\delta}}$ using the Boltzmann code CAMB \citep{Lewis:1999bs,2012JCAP...04..027H}. To include non-linear corrections we use HALOFIT \citep{Takahashi2012halofit}. The remaining power spectra used in this analysis are $P_{\rm{\delta I}}$, $P_{\rm{II}}$, $P_{\rm{gg}}$ and $P_{\rm{g \delta}}$. $P_{\rm{\delta I}}$ and $P_{\rm{II}}$ are the intrinsic alignment (IA) power spectra, which encode the tendency of galaxy shapes to point in the direction of a matter overdensity ($P_{\rm{\delta I}}$) or to have an intrinsic coherent alignment with other galaxy shapes ($P_{\rm{II}}$). $P_{\rm{gg}}$ summarises the clustering of galaxies, and $P_{\rm{g \delta}}$ the cross-correlations between galaxy position and gravitational shear. $P_{\rm{g \delta}}$ is linearly related to the galaxy-magnification power spectrum, which is the quantity of interest in this work. We employ a halo model formalism to calculate $P_{\rm{gg}}$ and $P_{\rm{g \delta}}$, while for the IA power spectra we use the empirical Non-linear Linear Alignment (NLA) model \citep{Hirata:2004gc,Bridle:2007ft}.

The halo model (e.g. \citealt{Cooray:2002dia}) assumes that dark matter clusters into dark matter halos and that all dark matter exists within dark matter halos. We define dark matter haloes as spheres of average density $\Delta \bar{\rho}_m$, with $\Delta = 200$ and $\bar{\rho}_m$ as the present day mean matter density of the Universe. Galaxies are then assumed to form within these dark matter halos, and hence the galaxy distribution traces the distribution of dark matter. The model relies on two ingredients, the underlying distribution of dark matter and how galaxies populate dark matter halos.  

The dark matter distribution is summarised by: the halo mass function, which gives the number density of dark matter halos with mass $M$ at redshift $z$; the halo bias function, which accounts for dark matter halos being biased tracers of the underlying dark matter distribution; and the halo density profile, which summarises how mass is distributed within dark matter halos. In this work we use the \cite{2010Tinker} functional forms for the halo mass function and halo bias function, and assume that the density of dark matter halos follows the Navarro-Frenk-White distribution \citep{1996NFWN}. To parametrise the concentration-mass relation that enters in the NFW profile, we follow \citet{Duffy2008}. We compute the halo mass function using the publicly available python package \textsc{hmf} \citep{Murray2013A&C.....3...23M,Murray:2020dcd}.

We summarise the second ingredient, how galaxies populate dark matter halos, using the conditional luminosity function (CLF) \citep{Yang2003MNRAS.339.1057Y,Cacciato2013MNRAS.430..767C,2013BoschCacciato}. The CLF gives the average number of galaxies with a luminosity $L$ between $L \pm \mathrm{d}L/2$ in a halo of mass $M$. It is divided into two parts:
\begin{equation}
  \Phi(L|M) = \Phi_{\rm{c}}(L|M) + \Phi_{\rm{s}}(L|M) \ ,
\end{equation}
where $\Phi_{\rm{c}}(L|M)$ is the CLF for central galaxies and $\Phi_{\rm{s}}(L|M)$ is the CLF for satellite galaxies. Central galaxies reside at the centre of dark matter halos and satellite galaxies orbit around them. Following the approach detailed in \cite{Cacciato2013MNRAS.430..767C} we take the CLF of central galaxies to be modelled by a lognormal distribution,
\begin{equation} \label{eq:centralLF}
  \Phi_{\rm{c}}(L|M)\mathrm{d}L = \frac{\log \mathrm{e}}{\sqrt{2\pi} \sigma_{\rm{c}}}\exp\bigg[-\frac{(\log L - \log L_{\rm{c}})^2}{2\sigma_{\rm{c}}^2}\bigg]\frac{\mathrm{d}L}{L} \ ,
\end{equation}
where $\sigma_{\rm{c}}$ represents the scatter in the log luminosity of central galaxies and $L_{\rm{c}}$ is parametrised as:
\begin{equation}
  L_{\rm{c}}(M) = L_0\frac{(M/M_{\rm{1}})^{\gamma_1}}{[1+(M/M_{\rm{1}})]^{\gamma_1-\gamma_2}} \ .
\end{equation}
$L_0 = 2^{\gamma_1-\gamma_2}L_{\rm{c}}(M_1)$ is a normalisation and $M_1$ is a characteristic mass scale. The CLF of satellite galaxies is modelled by a modified Schechter function,
\begin{equation} \label{eq:satelliteLF}
  \Phi_{\rm{s}}(L|M)\mathrm{d}L = \phi_{\rm{s}}^* \bigg(\frac{L}{L_{\rm{s}}^*}\bigg)^{\alpha_{\rm{s}}+1}\exp\bigg[-\bigg(\frac{L}{L^*_{\rm{s}}}\bigg)^2\bigg]\frac{\mathrm{d}L}{L} \ .
\end{equation}
where $\alpha_{\rm{s}}$ is the faint end slope of the satellite luminosity function. $\phi_{\rm{s}}^*$ is parametrised as:
\begin{equation}
  \log[\phi_{\rm{s}}^*(M)] = b_0 + b_1(\log M_{12}) + b_2(\log M_{12})^2,
\end{equation}
where $M_{12} = M/(10^{12}h^{-1}M_\odot)$ and $L_{\rm{s}}^*$ is parametrised as:
\begin{equation}
  L_{\rm{s}}^*(M) = 0.562 L_{\rm{c}}(M) \ .
\end{equation}
Both of the functional forms in eq. (\ref{eq:centralLF}) and (\ref{eq:satelliteLF}) are derived from the SDSS galaxy group catalog in \cite{Yang2008ApJ...676..248Y}. In total we have 9 free parameters in our CLF model: $\log M_1$, $\log L_0$, $\gamma_1$, $\gamma_2$, $\sigma_{\rm{c}}$, $\alpha_{\rm{s}}$, $b_0$, $b_1$ and $b_2$. We include all of these parameters in our Fisher matrix.

The Halo Occupation Distribution (HOD) can then be obtained as the integral of the CLF over the luminosity interval $[L_1, L_2]$:
\begin{equation} \label{eq:HOD_from_Phi}
  \langle N_{\rm{x}}|M\rangle = \int^{L_2}_{L_1} \Phi_{\rm{x}}(L|M) \mathrm{d}L \ ,
\end{equation}
where x can be c, s or g=c+s;
$\langle N_{\rm{c}}|M\rangle$ and $\langle N_{\rm{s}}|M\rangle$ are the average number of central and satellite galaxies in a halo of mass $M$ within the luminosity interval $[L_1,L_2]$.  Similarly, we can write $\bar{n}_{\rm{g}}$ as the average number density of galaxies across all halo masses in a given luminosity interval:
\begin{equation}
  \bar{n}_{\rm{g}}(z) = \int \langle N_{\rm{g}}|M\rangle n(M,z) \mathrm{d}M \ ,
\end{equation} 
where $n(M,z)$ is the halo mass function mentioned above.
To keep the notation compact, we have omitted the redshift dependence of the HOD: it arises as a consequence of the survey flux-limit: in this case, the luminosity limits $L_1$ and $L_2$ in eq. (\ref{eq:HOD_from_Phi}) depend on the specific redshift bin under consideration.

Once we have defined the HOD, we can calculate the 3D power spectra $P_{\rm{gg}}$ and $P_{\rm{g\delta}}$. First, the power spectra can be split into contributions from the one-halo (1h) and two-halo (2h) terms. The 1h term describes the clustering of galaxies on small scales within the same dark matter halo and the 2h term describes the clustering of galaxies on large scales between different halos. These contributions can then be split into the contributions from central c and satellite s galaxies, as with the CLF. This gives:
\begin{equation}
  \begin{split}
  & P_{\rm{gg}} = 2P_{\rm{cs}}^{\rm{1h}} + P_{\rm{ss}}^{\rm{1h}} + P_{\rm{cc}}^{\rm{2h}} + 2P_{\rm{cs}}^{\rm{2h}} + P_{\rm{ss}}^{\rm{2h}} \ , \\
  & P_{\rm{g\delta}} = P_{\rm{c\delta}}^{\rm{1h}} + P_{\rm{s\delta}}^{\rm{1h}} + P_{\rm{c\delta}}^{\rm{2h}} + P_{\rm{s\delta}}^{\rm{2h}} \ .
 \end{split}
\end{equation}
As shown in \cite{2013BoschCacciato} these contributions can be calculated using,
\begin{equation}
  \begin{split}
  P_{\rm{xy}}^{\rm{1h}}(k,z) = & \int \mathcal{H}_{\rm{x}}(k, M, z)\mathcal{H}_{\rm{y}}(k,M,z)n(M,z)\mathrm{d}M \ ,\\
  P_{\rm{xy}}^{\rm{2h}}(k,z) = & P^\mathrm{lin}_{\delta \delta} (k,z) \int \mathrm{d}M_1 \mathcal{H}_{\rm{x}}(k, M_1,z)n(M_1,z) b(M_1,z) \\
  & \times \int \mathrm{d}M_2 \mathcal{H}_{\rm{y}}(k,M_2,z)n(M_2,z) b(M_2,z) \ ,
 \end{split}
\end{equation}
where x and y can be c, s or $\delta$, and $b(M,z)$ is the halo bias.
The function $\mathcal{H}$ encodes the matter or galaxy contribution:
\begin{equation}
  \begin{split}
  & \mathcal{H}_{\rm{\delta}}(k,M,z) = \frac{M}{\bar{\rho}_{\rm{m}}}\tilde{u}_{\rm{h}}(k|M,z) \ , \\
  & \mathcal{H}_{\rm{c}}(k,M,z)=\mathcal{H}_{\rm{c}}(M,z) = \frac{\langle N_{\rm{c}}|M \rangle}{\bar{n}_{\rm{g}}(z)} \ , \\
  &\mathcal{H}_{\rm{s}}(k,M,z) = \frac{\langle N_{\rm{s}} | M \rangle}{\bar{n}_{\rm{g}}(z)}\tilde{u}_{\rm{s}}(k|M,z) \ .
  \end{split}
\end{equation}
where $\tilde{u}_{\rm{h}}$ is the Fourier transform of the normalised density distribution of dark matter in a halo of mass $M$ (mentioned above), and $\tilde{u}_{\rm{s}}$ is the normalised number density distribution of satellite galaxies in a halo of mass $M$. In his work, we assume satellites to follow the spatial distribution of the underlying dark matter, i.e. $\tilde{u}_{\rm{s}} \equiv \tilde{u}_{\rm{h}}$.

To calculate the 3D power spectra $P_{\rm{II}}$ and $P_{\rm{I\delta}}$ we employ the widely used NLA model. This model links the strength of the tidal field when a galaxy forms to the intrinsic ellipticity of the galaxy. This gives,
\begin{equation}
  \begin{split}
  & P_{\rm{\delta I}}(k,z)= -A_{\rm{IA}}C_1\rho_{\rm{c}} \frac{\Omega_{\rm{m}}}{D(z)} P_{\rm{\delta\delta}} \ , \\
  & P_{\rm{II}}(k,z)= \bigg( A_{\rm{IA}}C_1\rho_{\rm{c}} \frac{\Omega_{\rm{m}}}{D(z)} \bigg)^2 P_{\rm{\delta\delta}} \ ,
  \end{split} 
\end{equation} 
where $C_1$ is a normalisation constant, $\rho_{\rm{c}}$ the critical density of the Universe today and $D(z)$ the linear growth factor. We set $C_1 = 5 \times 10^{-14}M_{\odot}^{-1} h^{-2} \mathrm{Mpc}^3$ based on the IA amplitude measured at low redshifts using SuperCOSMOS \citep{Brown2002}, and $A_{\rm{IA}}$ captures the amplitude of the deviation from this reference case. We take $A_{\rm{IA}}$ as a free parameter in our Fisher matrix. The NLA model is sufficiently flexible for current studies but can be extended by including a redshift dependent parameter, or using a halo model formalism to calculate $P_{\rm{II}}$ and $P_{\rm{I\delta}}$ on small scales. Recently, \cite{Fortuna:2020vsz} explored these options and found that the IA signal in the one halo regime can be ignored at first order, and that including an extra redshift dependent parameter is possibly sufficient for LSST. Here we consider the simplest NLA model, but implementing more complex IA models could be a future extension of this work.

\section{Galaxy Luminosity Function} \label{sec:galaxy LF}

The second part of our Fisher matrix theory datavector, see section \ref{sec:fisher_analysis}, is the galaxy luminosity function. The galaxy luminosity function describes the distribution of luminosities in a galaxy sample, the number density of galaxies with a certain luminosity, and is often directly measured from a galaxy sample. As specified in \cite{Cacciato2013MNRAS.430..767C} the galaxy luminosity function at a given redshift $z$ can be calculated from the CLF detailed in section \ref{sec:3D power spectra}:
\begin{equation}
  \Phi(L,z) = \int dM \ \Phi(L|M)n(M,z) \ ,
\end{equation}
where $\Phi(L|M)$ is the CLF and $n(M,z)$ is the halo mass function (see section \ref{sec:3D power spectra}). In this analysis we work with a galaxy sample divided into redshift bins (labelled $i$ and $j$ previously) so we wish to compute the galaxy luminosity function for each redshift bin, 
\begin{equation} \label{eq:LF redshift binned}
  \Phi^i(L) = \int dz \ n^i(z) \Phi(L,z) \ ,
\end{equation}
where $\Phi^i(L)$ denotes the luminosity function of galaxies in redshift bin $i$, and $n^i(z)$ the normalised redshift distribution in bin $i$. We include a prediction for the galaxy luminosity function in each redshift bin in our theory datavector as it helps to constrain the 9 CLF parameters detailed in section \ref{sec:3D power spectra}, and hence is critical for obtaining information from the small scale clustering. The faint end slope of the number counts is also required to calculate the magnification 2D power spectra, see eq. (\ref{eq:alpha}).

\section{Survey Modelling} \label{sec:survey modelling}
We perform our Fisher forecast using the cosmological parameter estimation framework \textsc{CosmoSIS} \citep{Zuntz:2014csq}. To calculate the 3D power spectra detailed in section  \ref{sec:3D power spectra} we use our own halo model code, which has been tested against other halo model codes used in the literature.

\begin{figure*}
	\centering
	\begin{minipage}[b]{\columnwidth}
	\includegraphics[width=1.1\columnwidth]{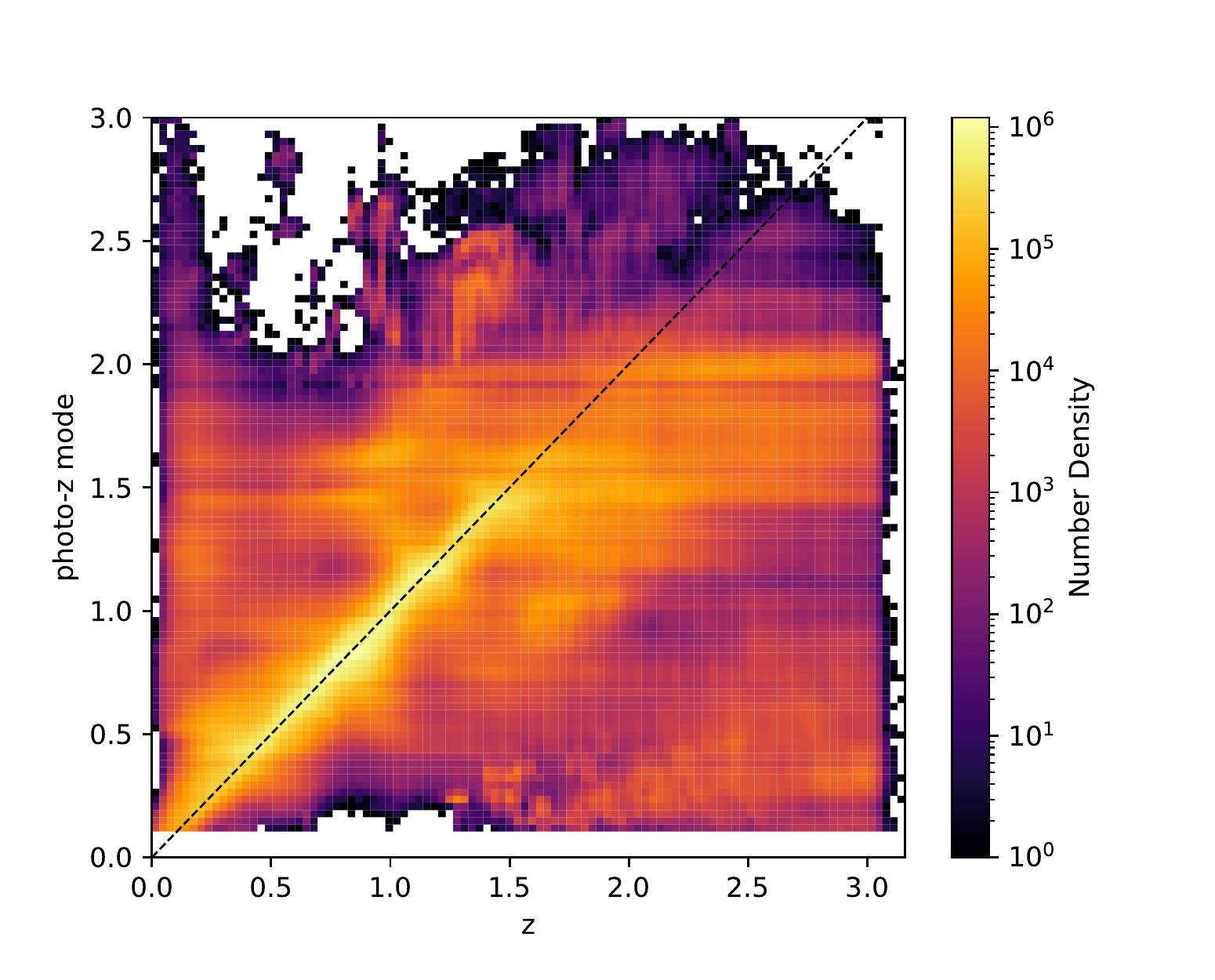}
	\end{minipage} \hfill 
	\begin{minipage}[b]{\columnwidth}
	\includegraphics[width=1.1\columnwidth]{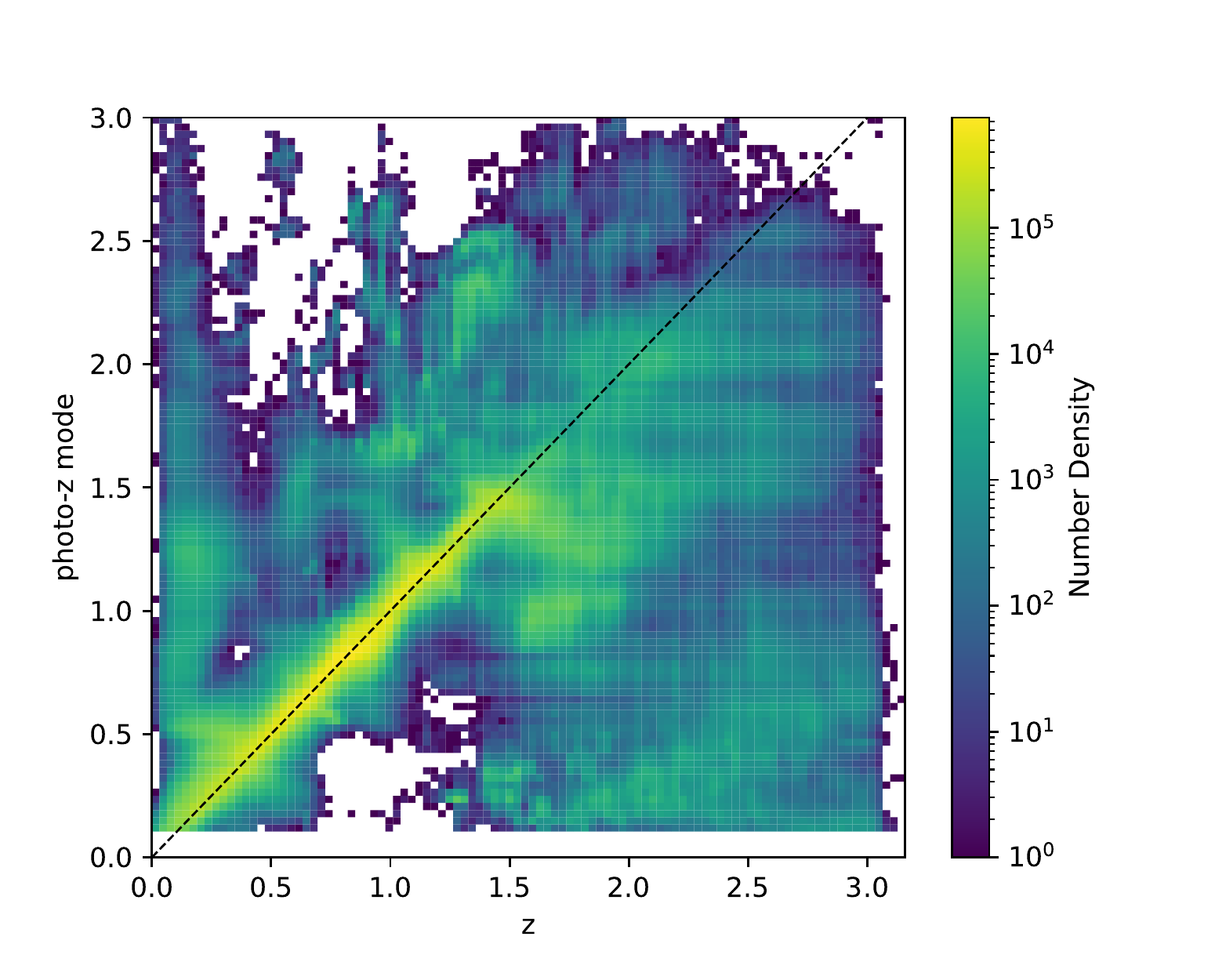}
	\end{minipage}
	\caption{Photometric redshift point estimate mode against true redshift. \textit{Left-hand panel:} number density sample \textit{n}-sample. \textit{Right-hand panel:} ellipticity sample $\epsilon$-sample}
  \label{fig:photoz_redshift}
\end{figure*}

In this analysis we define two mock LSST galaxy samples; an ellipticity sample $\epsilon$-sample and a number density sample \textit{n}-sample. We use a 440 square degree mock catalog from the LSST Dark Energy Science Collaboration (DESC) Data Challenge 2 (DC2) simulations (cosmoDC2 1.1.4; \citealt{Korytov:2019xus}). These simulations were designed to enable preliminary LSST DESC analyses, and the statistical distributions of galaxies have undergone a wide range of validation tests, for details see \cite{Korytov:2019xus,KovacsLSSTDESCDarkEnergyScience:2021lxq}. The catalog includes photometric redshifts for all galaxies with an i-band magnitude less than 26.5, up to redshift 3. The photometric redshifts were calculated using the template fitting code BPZ \citep{Benitez:1998br}. The \textit{n}-sample is defined as all galaxies in this mock catalog with an i-band magnitude less than 26.5 and photometric redshift greater than 0.1 and less than 2.0. We set an upper limit as the photometric redshifts begin to degrade significantly beyond 1.5, see Fig. \ref{fig:photoz_redshift}. The $\epsilon$-sample is defined as a subset of galaxies in \textit{n}-sample with $i < 25.3$. This corresponds to the LSST \textit{gold sample}, which will be used for weak lensing \citep{sciencebook}. We do not apply a separate signal-to-noise cut, but galaxies in the \textit{n}-sample have a signal-to-noise ratio $> 5$ and galaxies in the $\epsilon$-sample have a signal-to-noise ratio $> 20$. 

\subsection{Redshift distributions} \label{sec:nz dist}

\begin{figure*}
	\centering
	\begin{minipage}[b]{\columnwidth}
	\includegraphics[width=1.0\columnwidth]{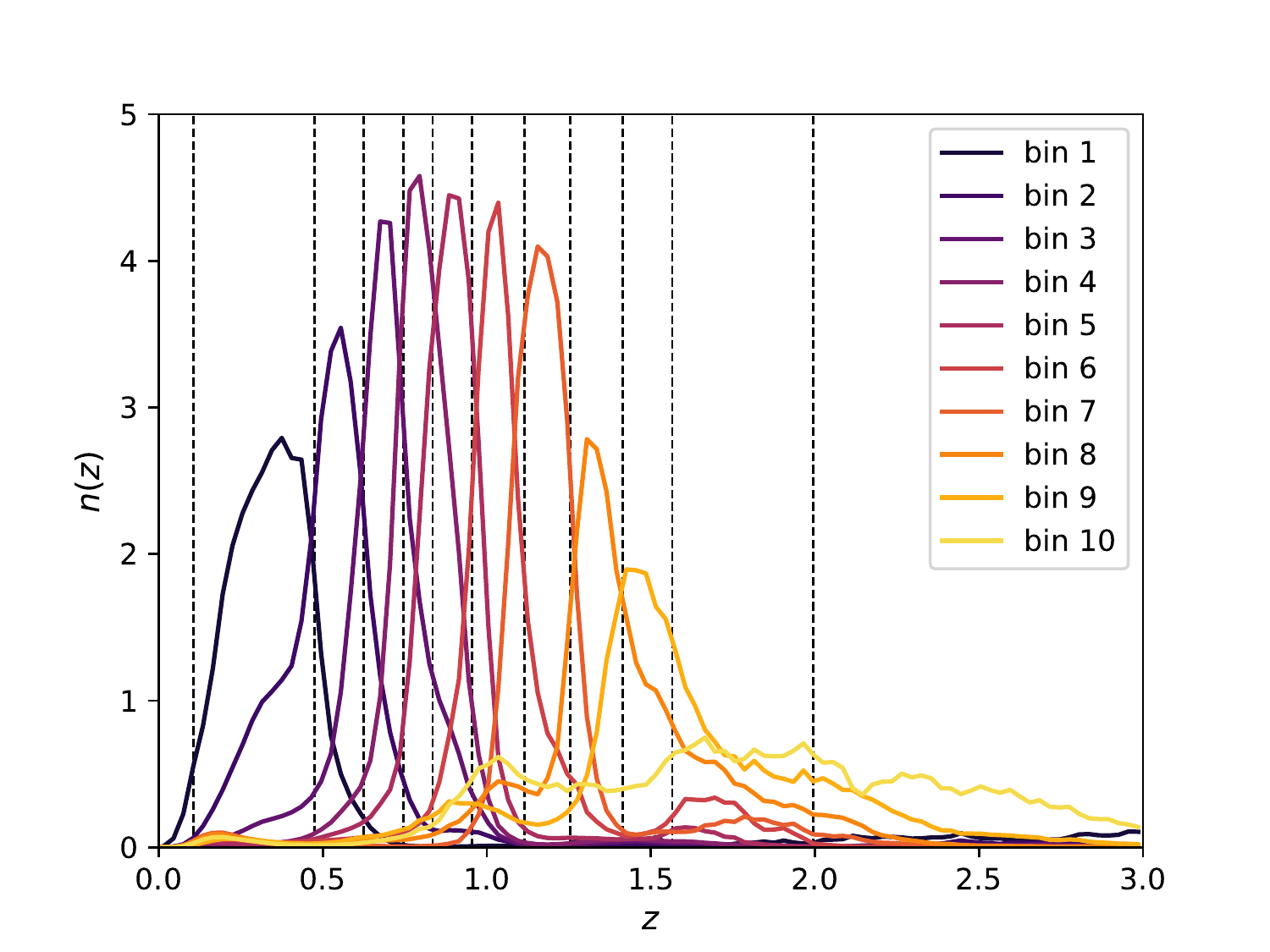}
	\end{minipage} \hfill 
	\begin{minipage}[b]{\columnwidth}
	\includegraphics[width=1.0\columnwidth]{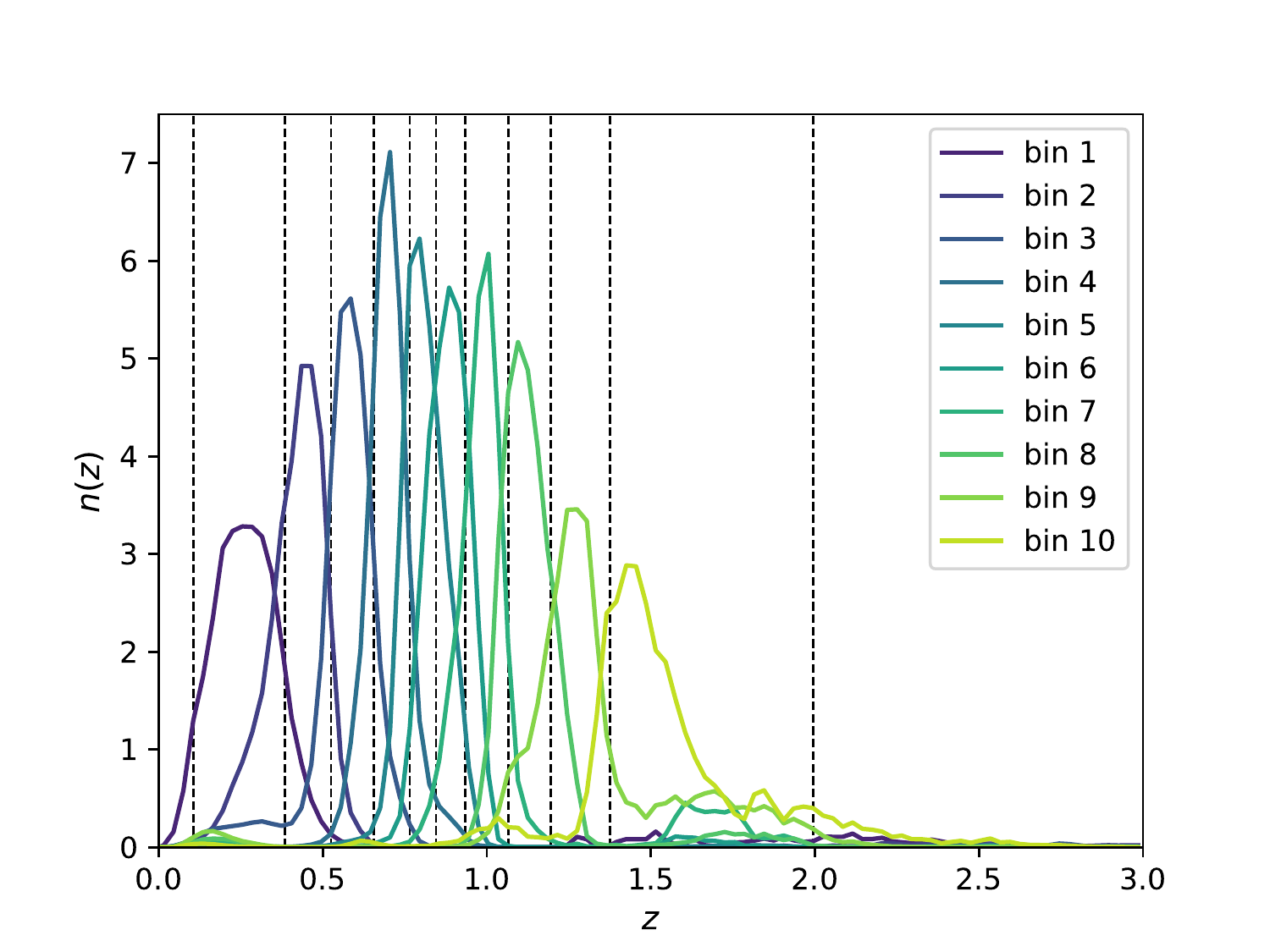}
	\end{minipage}
	\caption{Number density of galaxies as a function of true redshift for each photometric bin in the galaxy sample. The dashed lines indicate the photometric bin boundaries. \textit{Left-hand panel:} \textit{n}-sample. \textit{Right-hand panel:} $\epsilon$-sample}
  \label{fig:Nz}
\end{figure*}

To compute the 2D power spectra in eq. (\ref{eq:limber}) and the luminosity functions in eq. (\ref{eq:LF redshift binned}) we require the redshift distribution of galaxies in each photometric redshift bin. In this work we split both the galaxy samples, \textit{n}-sample and $\epsilon$-sample, into 10 tomographic redshift bins containing equal numbers of galaxies using their photometric redshifts. Figure \ref{fig:Nz} shows the resulting distribution of galaxies with redshift for each tomographic bin, as well as the tomographic bin boundaries. Figure \ref{fig:Nz} shows that the photometric redshifts are close to random for bin 10 of \textit{n}-sample, so our maximum photometric redshift cut of 2.0 is well justified. 

We compute the number density of galaxies in each tomographic bin to be $12.7\ \mathrm{arcmin}^{-2}$ for \textit{n}-sample and $4.9\ \mathrm{arcmin}^{-2}$ for $\epsilon$-sample. However, weak lensing shape measurements typically weight galaxies by the uncertainty or ability to calibrate the shape measurements, this would reduce the number density for $\epsilon$-sample, especially at high redshifts. The LSST science book estimates that the number density of galaxies in the gold sample will be 55 arcmin$^{-2}$, with the number density of galaxies useful for weak lensing  approximately 40 arcmin$^{-2}$ \citep{sciencebook,Chang2013MNRAS.434.2121C}. This means that our $\epsilon$-sample is slightly optimistic, with a galaxy number density of 49 arcmin$^{-2}$.

\subsection{Faint end number count slopes}

The key quantity in determining the amplitude of the fluctuation in the number density of galaxies as a result of weak lensing magnification is the faint end slope of the number counts $\alpha$. If $\alpha$ is equal to 1 there is no overall fluctuation but if $\alpha$ does not equal 1 there is either an increase or decrease in the number density of galaxies. $\alpha$ can be represented in terms of magnitudes as,
\begin{equation} \label{eq:alpha_mag}
\alpha(i_{\rm{mag}}) = 2.5 \frac{\mathrm{d}\log_{10}N(<i_{\rm{mag}})}{\mathrm{d}i_{\rm{mag}}} \ ,
\end{equation} 
where $i_{\rm{mag}}$ represents the $i$ band magnitude, and $N(<i_{\rm{mag}})$ the unlensed cumulative number density of galaxies with an $i$ band magnitude lower (brighter) than $i_{\rm{mag}}$ (e.g. \citealt{Duncan:2013haa}).

\begin{figure*}
  \centering
  \includegraphics[width=1.5\columnwidth]{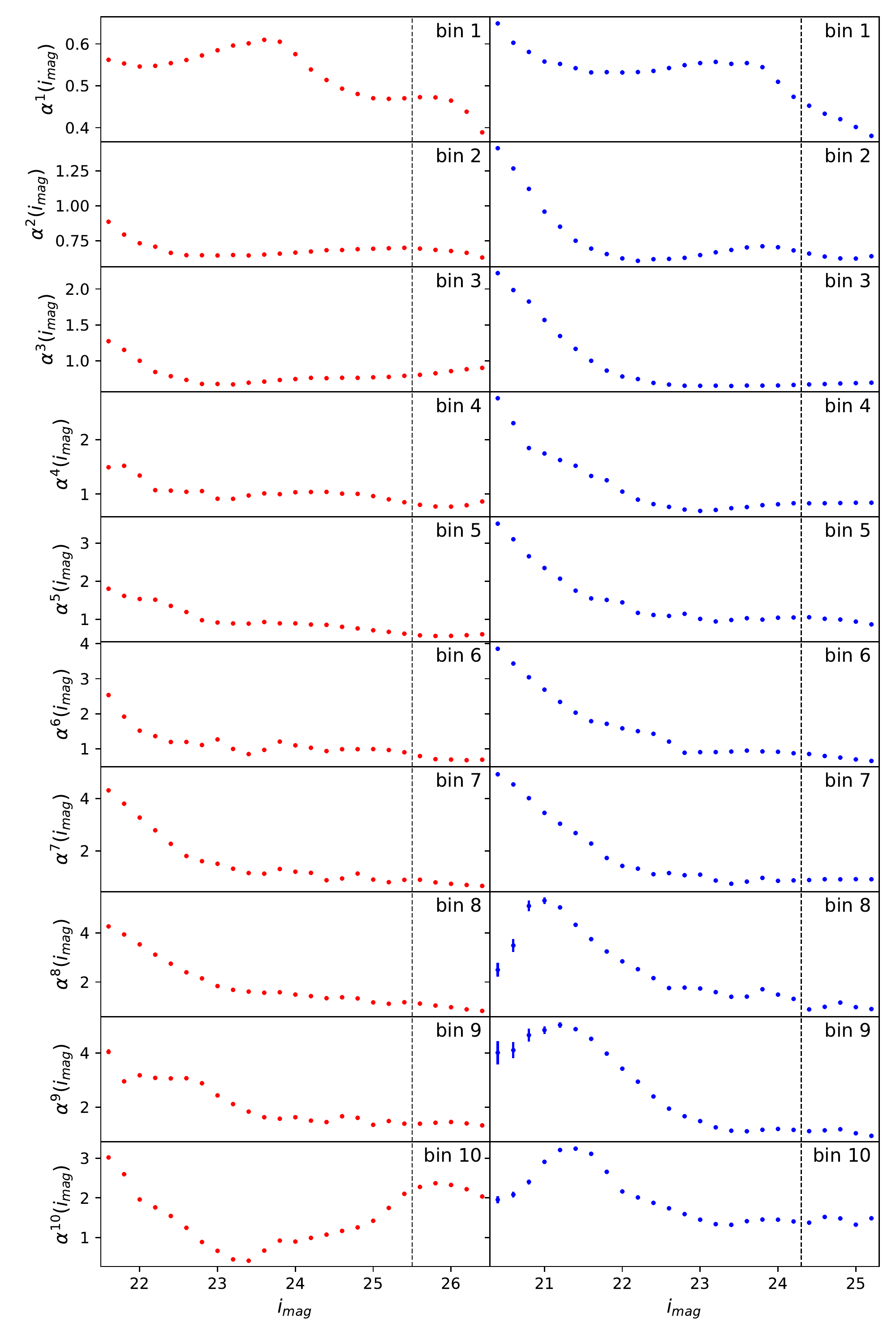}
  \caption{The faint end slope of the number counts $\alpha^i$ as a function of the limiting magnitude for each tomographic bin in \textit{n}-sample (red) and $\epsilon$-sample (blue). The $\alpha^i$ values used in this analysis were found by fitting the slope of the logarithmic cumulative number counts (see eq. (\ref{eq:alpha_mag})) between the vertical line and the right hand side of the figure.}
  \label{fig:alphas}
\end{figure*}

We measure the faint end slopes $\alpha$ from our LSST DC2 mock catalog. We compute a value $\alpha^j$ for each redshift bin $j$, in each mock sample. To compute $\alpha^j$ we vary the $i$ band magnitude in eq. (\ref{eq:alpha_mag}) and compute the cumulative number counts $N(>i_{\rm{mag}})$. We then fit the logarithm of $N(>i_{\rm{mag}})$ with a straight line, and use the slope to compute $\alpha^j$. Since we are only interested in the slope at the faint end (high magnitudes) we only fit $\log_{10}N(>i_{\rm{mag}})$ over the last magnitude before the sample magnitude limit; 25.5-26.5 for \textit{n}-sample, and 24.3-25.3 for $\epsilon$-sample. Figure \ref{fig:alphas} shows that in general this lower fit limit (marked by a dotted line) captures the value of $\alpha^j$ at the faint end of the sample. Increasing the lower fit limit has little effect on the value of $\alpha^j$ obtained, whereas decreasing the fit limit in general gives a higher value of $\alpha^j$. 

Table \ref{tab:alpha_values} shows the $\alpha^j$ values obtained for each sample and their associated uncertainties. The uncertainties come from the uncertainty on the slope coefficient of the least-squares straight line fit detailed above, since they were found to be much larger than the uncertainties on the values of the cumulative number counts $N(>i_{\rm{mag}})$ due to the large number of galaxies in each sample. The uncertainties are very small, and would become even smaller when using the full 18000 square degree LSST area instead of a 440 square degree mock catalog. We therefore consider the $\alpha^j$ parameters as fixed in our forecast, but note that they can be difficult to measure accurately from real data due to the presence of systematics and selection effects (see conclusions for further discussion).

We can compare the $\alpha$ values in Table \ref{tab:alpha_values} to those found in \cite{Duncan:2013haa} for the Canada--France--Hawaii Lensing Survey (CFHTLenS). In both cases $\alpha^j$ generally increases with redshift. CFHTLenS reaches an $\alpha^j$ value of approximately 1 at its $i$ band magnitude limit of 24.7, for its highest redshift bin between 1.02 and 1.3. This roughly corresponds to $\alpha^7$ and $\alpha^8$ in $\epsilon$-sample, where the magnitude limit of 24.7 is included in the $\alpha^j$ fit. Table \ref{tab:alpha_values} shows that our $\alpha^7$ and $\alpha^8$ values for $\epsilon$-sample are consistent with CFHTLenS.
\begin{table}
  \centering
  \caption{Faint end number count slopes $\alpha^j$ for each redshift bin $j$ in \textit{n}-sample and $\epsilon$-sample, with their associated 1$\sigma$ uncertainties.}
  \begin{tabular}{ll@{\hskip 1cm}ll}
  \multicolumn{2}{l}{\textbf{n-sample}} & \multicolumn{2}{l}{\textbf{\bm{$\epsilon$}-sample}} \\
  \hline \\
  $\alpha^1$ & $0.445\pm0.005$ & $\alpha^1$ & $0.412\pm0.005$  \\
  $\alpha^2$ & $0.663\pm0.006$ & $\alpha^2$ & $0.624\pm0.004$  \\
  $\alpha^3$ & $0.848\pm0.006$ & $\alpha^3$ & $0.677\pm0.004$   \\
  $\alpha^4$ & $0.781\pm0.005$ & $\alpha^4$ & $0.825\pm0.006$   \\
  $\alpha^5$ & $0.573\pm0.004$ & $\alpha^5$ & $0.97\pm0.01$  \\
  $\alpha^6$ & $0.694\pm0.006$ & $\alpha^6$ & $0.74\pm0.01$  \\
  $\alpha^7$ & $0.74\pm0.01$   & $\alpha^7$ & $0.895\pm0.006$  \\
  $\alpha^8$ & $0.95\pm0.02$   & $\alpha^8$ & $0.99\pm0.01$  \\
  $\alpha^9$ & $1.39\pm0.01$   & $\alpha^9$ & $1.08\pm0.01$  \\
  $\alpha^{10}$ & $2.24\pm0.02$ & $\alpha^{10}$ & $1.42\pm0.01$                     
  \end{tabular}\label{tab:alpha_values}
  \end{table}

\subsection{Systematics} \label{sec:systematics}

We include a number of systematics in our analysis using nuisance parameters. For the fiducial values of these parameters and their associated priors please see Table \ref{tab:fiducial values}. To apply a Gaussian prior to a particular parameter in a Fisher matrix, one simply adds $1/\sigma^2_{\rm{prior}}$ to the diagonal element associated with the parameter \citep{Coe:2009xf}. In conceptual terms, the priors on the Fisher matrix parameters can be summarized by a diagonal covariance matrix with elements $\sigma^2_{\rm{prior}}$. This covariance matrix can then be inverted into a prior Fisher matrix, giving $1/\sigma^2_{\rm{prior}}$ diagonal elements, and added to the experimental Fisher matrix.

\subsubsection{Shear multiplicative bias}
Systematic uncertainties in the measuring and averaging of galaxy shapes can result in a multiplicative scaling of the observed shear. These systematic effects include: noisy galaxy images, the applicability of the model used to describe the light profile of galaxies, the details of the galaxy morphology and selection biases (e.g. \citealt{heymans10.1111/j.1365-2966.2006.10198.x, Mandelbaum:2017ctf,Zuntz:2017pso,Kannawadi:2018moi}). We parametrise this multiplicative scaling using one parameter $m^i$ per redshift bin (10 parameters in total), which scale the cosmic shear and galaxy-galaxy lensing power spectra as:
\begin{equation}
\begin{split}
& C_{\rm{\epsilon\epsilon}}^{ij}(l) \rightarrow (1+m^i)(1+m^j) C_{\rm{\epsilon\epsilon}}^{ij}(l) \ , \\
& C_{\rm{n\epsilon}}^{ij}(l) \rightarrow (1+m^j)C_{\rm{n\epsilon}}^{ij}(l) \ .\\
\end{split}
\end{equation}
We impose Gaussian priors on these multiplicative parameters, which are guided by the LSST DESC science requirements \citep{Mandelbaum:2018ouv}. These science requirements forecast the uncertainties LSST will need to achieve in order to meet their main objectives of significantly improving the constraints on the dark energy parameters $w_0$ and $w_a$, compared to previous dark energy experiments, and obtaining dark energy constraints where the total calibratable systematic uncertainty is less than the marginalised statistical uncertainty. For the case of shear multiplicative bias the requirement is that the `systematic uncertainty in the redshift-dependent shear calibration' should not exceed 0.003 by year 10. We therefore apply a Gaussian prior centred on zero with a standard deviation of 0.003 to each of our shear multiplicative bias parameters.

\subsubsection{Clustering Multiplicative Bias}
We parametrise uncertainties in the number count measurement using a similar approach to that for shear. Systematics which affect the number density of galaxies include: galactic dust obscuring background galaxies, variable survey depth impacting the number of sources promoted across the flux limit by magnification, and stars contaminating the galaxy sample \citep{hildebrandt10.1093/mnras/stv2575, Thiele:2019fcu}. Usually these effects would be partially absorbed by the galaxy bias (CLF) parameters, however since we include the galaxy luminosity function in our analysis the CLF parameters will be tightly constrained. We therefore felt it was important to include this multiplicative bias parameterisation for clustering as well as shear.

Analogous to shear multiplicative bias, the observed clustering power spectra are scaled by a multiplicative factor as,
\begin{equation}
  \begin{split}
  & C_{\mathrm{nn}}^{ij}(l) \rightarrow (1+ m_{\mathrm{eff}} ^{i}) (1+ m_{\mathrm{eff}}^{j})C_{\mathrm{nn}}^{ij}(l) \ , \\
  & C_{\mathrm{n\epsilon}}^{ij}(l) \rightarrow (1+ m_{\mathrm{eff}} ^{i}) C_{\mathrm{n\epsilon}}^{ij}(l) \ . 
  \end{split}
  \end{equation}
However, since most systematics decrease with signal to noise ratio, we assume $m^i_{\mathrm{eff}}$ has a power law dependence on the signal to noise of galaxies in redshift bin $i$. This enables us to reduce the number of clustering multiplicative bias parameters from ten parameters (one $m^i_{\mathrm{eff}}$ per redshift bin) to two parameters $a_{\rm{m}}$ and $b_{\rm{m}}$. $m^i_{\mathrm{eff}}$ is given in terms of $a_{\rm{m}}$ and $b_{\rm{m}}$ by,
\begin{equation}
  \begin{split}
  m^i_{\mathrm{eff}} & = m_{\mathrm{step}} - m_{\mathrm{fid}} \\
  & = \frac{1}{N_i} \Bigg[ a_{\rm{m}} \sum_{n=1}^{N_i} \Big(\frac{S}{N}\Big)_n^{b_{\rm{m}}} - a_{\mathrm{fid}} \sum_{n=1}^{N_i} \Big(\frac{S}{N}\Big)_n^{b_{\mathrm{fid}}} \Bigg] \ ,
  \end{split}
  \end{equation}
where $N_i$ is the number of galaxies in tomographic bin $i$, the sum is over the signal-to-noise ratio $S/N$ of all galaxies in tomographic bin $i$, $a_{\mathrm{fid}}$ is the fiducial value of $a_{\rm{m}}$ and $b_{\mathrm{fid}}$ is the fiducial value of $b_{\rm{m}}$. We introduce the $m_{\mathrm{fid}}$ term because if $m_{\mathrm{eff}} = m_{\mathrm{step}}$, $b_{\rm{m}}$ becomes unconstrained when $a_{\rm{m}}$ is equal to zero, which breaks the Gaussian Likelihood assumption in the Fisher matrix prediction.

We compute the signal to noise ratio for each galaxy in our samples from the error on the i band apparent magnitude. Using the signal to noise of every galaxy in this bias calculation is computationally expensive, since the total number of galaxies in \textit{n-sample} and $\epsilon$\textit{-sample} is of order $10^7$ and $10^8$. We therefore use a randomly selected $1\%$ subsample of galaxies in this calculation. This subsample is representative of the full galaxy sample, but prevents our bias calculation from being prohibitively slow.

\subsubsection{Photometric redshift uncertainties}

We model uncertainties in the redshift distributions shown in figure \ref{fig:Nz} by introducing shift factors $\Delta^i$ \citep{Bonnett:2015pww}. $\Delta^i$ simply shifts the redshift distribution in bin $i$ so,
\begin{equation}
n^i(z) \rightarrow n^i(z-\Delta^i).
\end{equation}
Since we have two redshift distributions, one for $\epsilon$-sample and one for \textit{n}-sample, each divided into 10 bins this results in 20 shift parameters $\Delta^i$. These parameters are likely to be correlated, so we are making a conservative choice by allowing 20 separate shift parameters, which may somewhat weaken our final constraints. We impose Gaussian priors on each of these shift parameters, once again guided by the LSST DESC science requirements \citep{Mandelbaum:2018ouv}. The prior is centred on zero, with a standard deviation of 0.003 for the \textit{n}-sample parameters and of 0.001 for the $\epsilon$-sample parameters.

A future extension of this work could be to include other modes of redshift uncertainty, such as a change in the width or to the high redshifts tails, as in \cite{Nicola:2019yiw}. These may be particularly interesting for magnification, as they change the level of overlap between different redshift bins.

\subsection{Covariances}

In this forecast we consider two component Fisher matrices. The Fisher matrix for the weak lensing observables and the Fisher matrix for the galaxy luminosity function (see section \ref{sec:fisher_analysis}). We therefore require two covariances: the weak lensing observables covariance and the galaxy luminosity function covariance.

\subsubsection{Weak lensing observables covariance} \label{sec:2pt cov}

We compute a Gaussian covariance for the observable weak lensing power spectra ($C_{\epsilon\epsilon}$, $C_{\mathrm{nn}}$, $C_{\mathrm{n\epsilon}}$) using \textsc{CosmoSIS}. The covariance between two power spectra is given by,
\begin{equation} \label{eq:weak_lensing_cov}
  \mathrm{Cov}\big[C^{ij}(\ell), \ C^{kl}(\ell')\big] =\delta_{\ell\ell'}\frac{2\pi}{A\ell\Delta \ell} \big[\bar{C}^{ik}(\ell)\bar{C}^{jl}(\ell)+\bar{C}^{il}(\ell)\bar{C}^{jk}(\ell)\big] \ ,
\end{equation}
where $ijkl$ denote redshift bins, $\delta_{\ell\ell'}$ is the Kronecker delta, $A$ is the survey area and $\Delta \ell$ the size of the angular frequency $\ell$ bin \citep{Joachimi:2007xd, Joachimi:2010}. We do not include the non-gaussian contributions to the covariance since their effect is small, and unlikely to impact our final results \citep{Barreira:2018jgd}. To account for the random terms in equations \ref{eq:epsilon_contirbutions} and \ref{eq:n_constributions} we define,
\begin{equation}
  \bar{C}^{ij}(\ell) = C^{ij}(\ell) + N^{ij} \ ,
\end{equation}
where $N^{ij}$ is the shot or shape noise contribution. In the case of $C_{\rm{\epsilon\epsilon}}$,
\begin{equation} \label{eq:sigma_e}
  N^{ij} = \delta_{ij} \frac{\sigma^2_{\epsilon}}{2\bar{n}^i} \ ,
\end{equation}
in the case of $C_{\rm{nn}}$,
\begin{equation}
  N^{ij} = \delta_{ij} \frac{1}{\bar{n}^i} \ ,  
\end{equation}
and in the case of $C_{\rm{n\epsilon}}$, $N^{ij}=0$. Where $\sigma_{\epsilon}$ is the total intrinsic ellipticity dispersion, and $\bar{n}^i$ is the average number density of galaxies in redshift bin $i$ \citep{Bartelmann:1999yn}. We compute the power spectra covariance for 20 log-spaced angular frequency $l$ bins from $l_{\rm{min}} = 30$, to avoid inaccuracies in the Limber approximation, to $l_{\rm{max}} = 3000$, to avoid the very non-linear regime.

\subsubsection{Galaxy luminosity function covariance}

We compute the galaxy luminosity function covariance by measuring the galaxy luminosity functions of our mock LSST galaxy samples and then computing a bootstrap covariance. Since Fisher forecasts do not require a datavector, only a covariance, we only use the measured luminosity functions to compute the covariance and model the galaxy luminosity function in the forecast using the CLF formalism (see section \ref{sec:galaxy LF}).

To measure the luminosity functions for the \textit{n}-sample and $\epsilon$-sample we begin by computing the luminosity of each galaxy from its rest-frame absolute magnitude in the $i$ band. We then divide our sample into the 10 tomographic bins described above and scale the luminosity function for each bin $j$ by the volume of bin $j$, to convert the histogram to a number density. When calculating the bin volume we assume that the galaxies do not scatter beyond the tomographic bin boundaries. This is an approximation, which figure \ref{fig:Nz} shows, is becoming problematic for bin 10.

Ideally, we would use the full range of galaxy luminosities to compute our bootstrap covariance. However in order to use the low luminosity region we would need to correct our galaxy samples to be volume complete, for example through the $1/V_{\mathrm{max}}$ method \citep{1968schmidtApJ...151..393S,1976FeltenApJ...207..700F,Cole2011MNRAS.416..739C}. High luminosity objects can be observed across the full volume of the survey, but low luminosity objects can only be observed at smaller distances. This introduces a bias referred to as Malmquist bias, and we therefore only want to include galaxies that can be observed across the whole volume of the survey. For the purposes of this work we deemed it sufficient to simply cut out the low luminosity galaxies to make the sample volume limited, since this is still a significant step forward compared to previous analyses. For details of how we determine the volume complete cut see appendix \ref{sec:vol_complete_cut}.

We then compute a bootstrap covariance for our measured galaxy luminosity functions. First, we sample our dataset with replacement 100 times and compute the associated datavectors. We then assume that each luminosity bin in each tomographic bin is independent (each of our datapoints is independent) and calculate the variance of these 100 samples. This gives us a diagonal covariance. The variance of the 100 samples is in general small, due to the very large numbers of galaxies in each sample.

\subsection{Fiducial values} \label{sec:CLF fiducial values}

The Fisher matrix gives the curvature of the log-Likelihood function around its peak. It does not find the location of the peak, this is defined with a set of fiducial values (shown in Table \ref{tab:fiducial values}). The set of parameters required to calculate the 3D power spectra in section \ref{sec:3D power spectra} are the cosmological parameters and the CLF parameters. In this work we consider the constraints on a flat $\Lambda$CDM cosmology, and vary the cosmological parameters; $\Omega_{\rm{m}}$ the matter density, $h_0$ the hubble parameter, $\Omega_{\rm{b}}$ the baryon density, $n_{\rm{s}}$ the scalar spectral index, $A_{\rm{s}}/10^{-9}$ the amplitude of primordial fluctuations, and $w$ and $w_{\rm{a}}$ the dark energy equation of state parameters. We take their fiducial values from the input values used to generate the simulation for the LSST DESC mock catalog, or from the values obtained by the \textit{Planck} satellite \citep{Planck:2018vyg}. 
\begin{table}
  \centering
  \caption{Fiducial values and priors for the model parameters used to compute the fisher matrices in this work. Flat priors do not contribute to the Fisher matrix so we simply specify flat, and do not include bounds.}
  \begin{tabular}{lll}
  \textbf{Parameter} & \textbf{Fiducial Value} & \textbf{Prior}  \\
  \hline
  \hline
            & \textbf{Survey}&  \\
  Area      & $18000 \ \mathrm{deg^2}$ & fixed \\
  $\sigma_e$      & $0.35$ & fixed \\
  \hline
            & \textbf{Cosmology} &  \\
       $\Omega_{\rm{m}}$     &  $0.265$ & flat \\
            $h_0$     &  $0.71$ & flat \\
            $\Omega_{\rm{b}}$     & $0.0448$ & flat \\
            $n_{\rm{s}}$     & $0.963$ & flat \\
            $A_{\rm{s}}/10^{-9}$     &  $2.1$ & flat \\
            $w$     &  $-1.0$ & flat \\
            $w_{\rm{a}}$     & $0.0$ & flat \\
            $\Omega_{\rm{k}}$ & $0.0$ & fixed \\
  \hline
            & \textbf{CLF}&  \\
            $\log(M_1)$     & $11.24$ & flat \\
            $\log(L_0)$     & $9.95$ & flat \\
            $\gamma_1$     & $3.18$ & flat \\
            $\gamma_2$     & $0.245$ & flat \\
            $\sigma_c$     & $0.157$ & flat \\
            $\alpha_{\rm{s}}$     & $-1.18$ & flat \\
            $b_0$     & $-1.17$ & flat \\
            $b_1$     & $1.53$ & flat \\
            $b_2$     & $-0.217$ & flat \\
  \hline
            & \textbf{Intrinsic Alignments}& \\
            $A_{\rm{IA}}$ & $1.0$ & flat \\
  \hline
            & \textbf{\textit{n}-sample Photo-z} &  \\
            $\Delta_{\rm{n}}^i$ & $0.0$ & Gauss(0.0, 0.003) \\
  \hline
            & \textbf{$\epsilon$-sample Photo-z} &  \\
            $\Delta_{\rm{\epsilon}}^i$ & $0.0$ & Gauss(0.0, 0.001) \\
  \hline
            & \textbf{Shear Bias} &  \\
            $m^i$ & $0.0$ &  Gauss(0.0, 0.003)\\
  \hline
            & \textbf{Clustering Bias} & \\
            $a_{\rm{m}}$ & $0.001$ & flat\\
            $b_{\rm{m}}$ & $0.0$ & flat
  \end{tabular}
  \label{tab:fiducial values}
  \end{table} 

We also vary the full set of CLF parameters $\log M_1$, $\log L_0$, $\gamma_1$, $\gamma_2$, $\sigma_{\rm{c}}$, $\alpha_{\rm{s}}$, $b_0$, $b_1$ and $b_2$, detailed in section \ref{sec:3D power spectra}. Here we use the fiducial values found for SDSS by \cite{Cacciato2013MNRAS.430..767C}, which have been shown to also be applicable to higher redshift surveys \citep{2014caccaitoedoMNRAS.437..377C,2016edoA&A...586A..43V}.

\section{Results} \label{sec:results}

\subsection{Clustering} \label{sec:clustering results}
\begin{figure*}
  \centering
  \includegraphics[width=1.5\columnwidth]{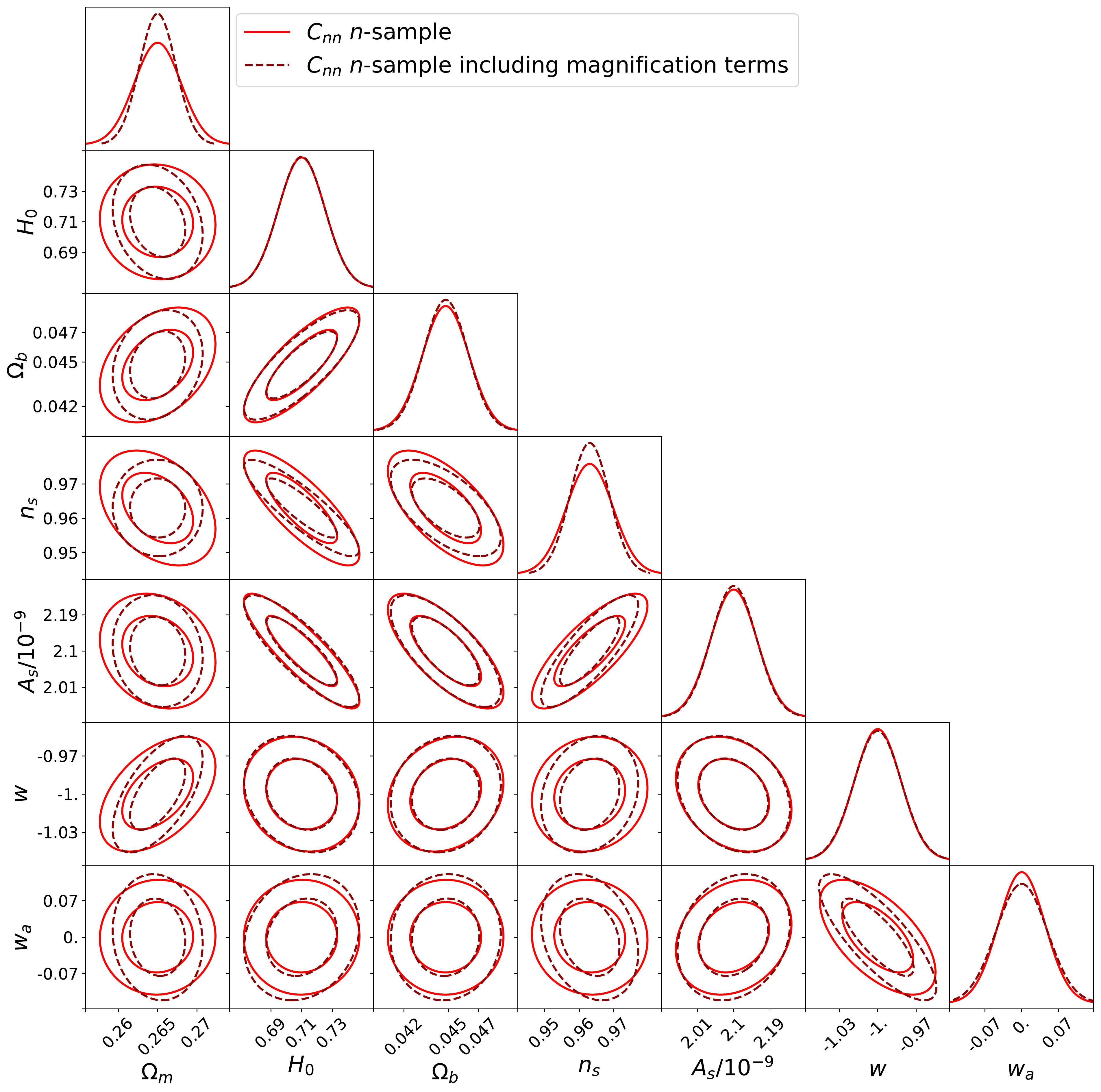}
  \caption{Constraints on the cosmological parameters used in this analysis from $C_{\mathrm{nn}}$ and $C_{\mathrm{nn}}$ including magnification terms for \textit{n}-sample. Including magnification has only a small impact on the constraints.}
  \label{fig:Cnn_nsample_cos_params}
\end{figure*}

Figure \ref{fig:Cnn_nsample_cos_params} shows the forecast constraints on the cosmological parameters from $C_{\mathrm{nn}}$ with and without including magnification terms for \textit{n}-sample. In the case of including magnification the observable is $C_{\mathrm{nn}} = C_{\mathrm{gg}} + C_{\mathrm{gm}} + C_{\mathrm{mm}}$ instead of $C_{\mathrm{nn}} = C_{\mathrm{gg}}$. Including magnification generally has a small impact on the cosmological parameter constraints. The greatest change is the 1$\sigma$ constraint on $\Omega_m$, which is improved by a factor of 1.3 from 0.003 to 0.0023.

The forecast constraints on the cosmological parameters from $C_{\mathrm{nn}}$ and $C_{\mathrm{nn}}$ including magnification terms for $\epsilon$-sample show that the impact of magnification is reduced compared to the \textit{n}-sample. The 1$\sigma$ constraint on $\Omega_m$ is only improved by a factor of 1.03 from 0.0032 to 0.0031, instead of a factor of 1.3 with the \textit{n}-sample. This shows that including magnification has a greater impact for deeper samples.

\begin{figure*}
  \centering
  \includegraphics[width=1.5\columnwidth]{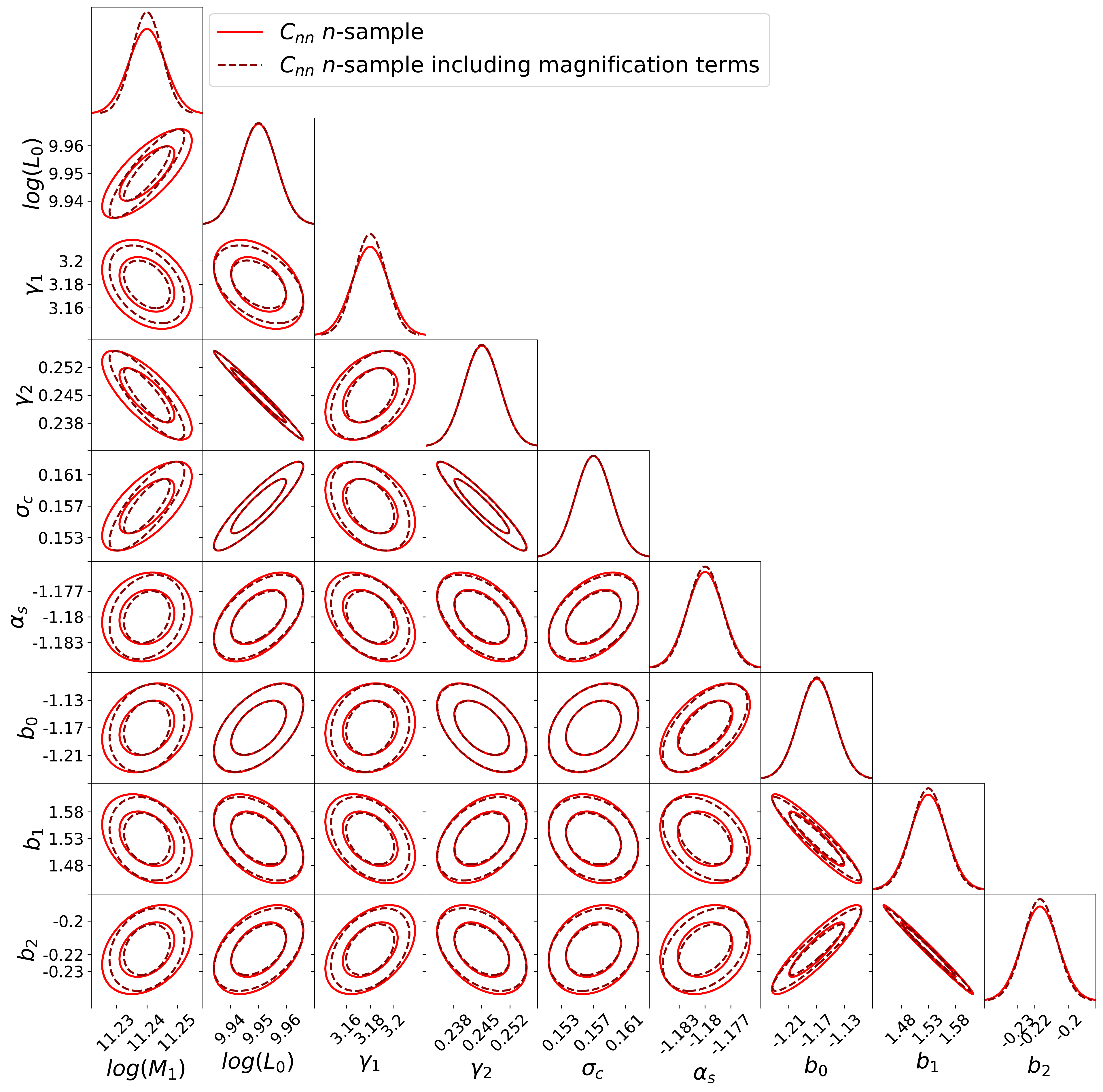}
  \caption{Constraints on the Conditional Luminosity Function (CLF) parameters used in this analysis from $C_{\mathrm{nn}}$ and $C_{\mathrm{nn}}$ including magnification terms for \textit{n}-sample. Including magnification has little effect on the constraints, since they are driven by the galaxy luminosity function not the weak lensing or clustering observables.}
  \label{fig:Cnn_nsample_hod_params}
\end{figure*}

Figure \ref{fig:Cnn_nsample_hod_params} shows the forecast constraints on the CLF parameters from $C_{\mathrm{nn}}$ with and without including magnification terms for the \textit{n}-sample. Including magnification has little effect on the constraints on the CLF parameters. This is expected because the CLF constraints are predominantly determined by the galaxy luminosity function. We focus on the cosmological and CLF parameters, instead of presenting the full 28 parameter space, for clarity. The steps taken to ensure the stability of our Fisher matrix are detailed in appendix \ref{sec: Fisher matrix stability}.

A useful measure of the constraining power of an analysis is the Figure of Merit (FoM) defined as,
\begin{equation}
  {\rm{FoM} = det([F^{-1}]}_{q})^{\frac{1}{N_{q}}} \ ,
\end{equation}
where ${\rm{[F^{-1}]}}_q$ is the inverse Fisher matrix for the set of parameters $q$ and $N_q$ is the number of parameters $q$ in the set. In this work we define $q$ as the full set of cosmological parameters, so the FoM represents the power of the constraints on the cosmological parameters. It is also common to define a Dark Energy FoM where $q=\{w, w_a\}$ \citep{Albrecht:2006um}.

When magnification is included in the clustering analysis for the \textit{n}-sample the FoM is increased by a factor of 1.45. However, when magnification is included in the clustering analysis for $\epsilon$-sample (the LSST gold sample) the FoM is increased by a factor of 1.08. This mirrors the conclusions from looking at the parameter constraints on  $\Omega_m$ -- magnification is more beneficial for deeper samples with greater numbers of low signal-to-noise ratio galaxies. Interestingly, there is no increase in the FoM for clustering without magnification when using the \textit{n}-sample instead of $\epsilon$-sample. This implies that it is more beneficial to have a smaller sample of high signal-to-noise objects than a larger sample including lower signal-to-noise objects. This is likely due to the additional fainter objects having poorer photometric redshifts and therefore largely contributing to the tails of the redshift distribution. Looking back at figure \ref{fig:Nz} we can see that the redshift distribution for the $\epsilon$-sample is much cleaner.

\subsection{Shear calibration} \label{sec:shear calibration}

The previous section showed that including weak lensing magnification only has a small effect on the cosmological parameter constraints from an LSST-like angular galaxy clustering analysis. In a combined clustering and cosmic shear analysis the impact of magnification on the cosmological parameter constraints can only be reduced. This is because magnification predominantly contributes to the clustering signal and provides very similar information to shear. We therefore focus on the effect of magnification on the shear multiplicative bias parameters.

We examine the impact of including magnification on the shear multiplicative bias parameters for a combined LSST clustering $C_{\mathrm{nn}}$ and shear $C_{\rm{\epsilon\epsilon}}$ analysis, where the analyses occur on separate patches of sky so the $C_{\rm{n\epsilon}}$ term is negligible. We are therefore investigating whether the improved cosmological constraints from magnification translate into an improved calibration.  

\begin{figure*}
  \centering
  \includegraphics[width=1.5\columnwidth]{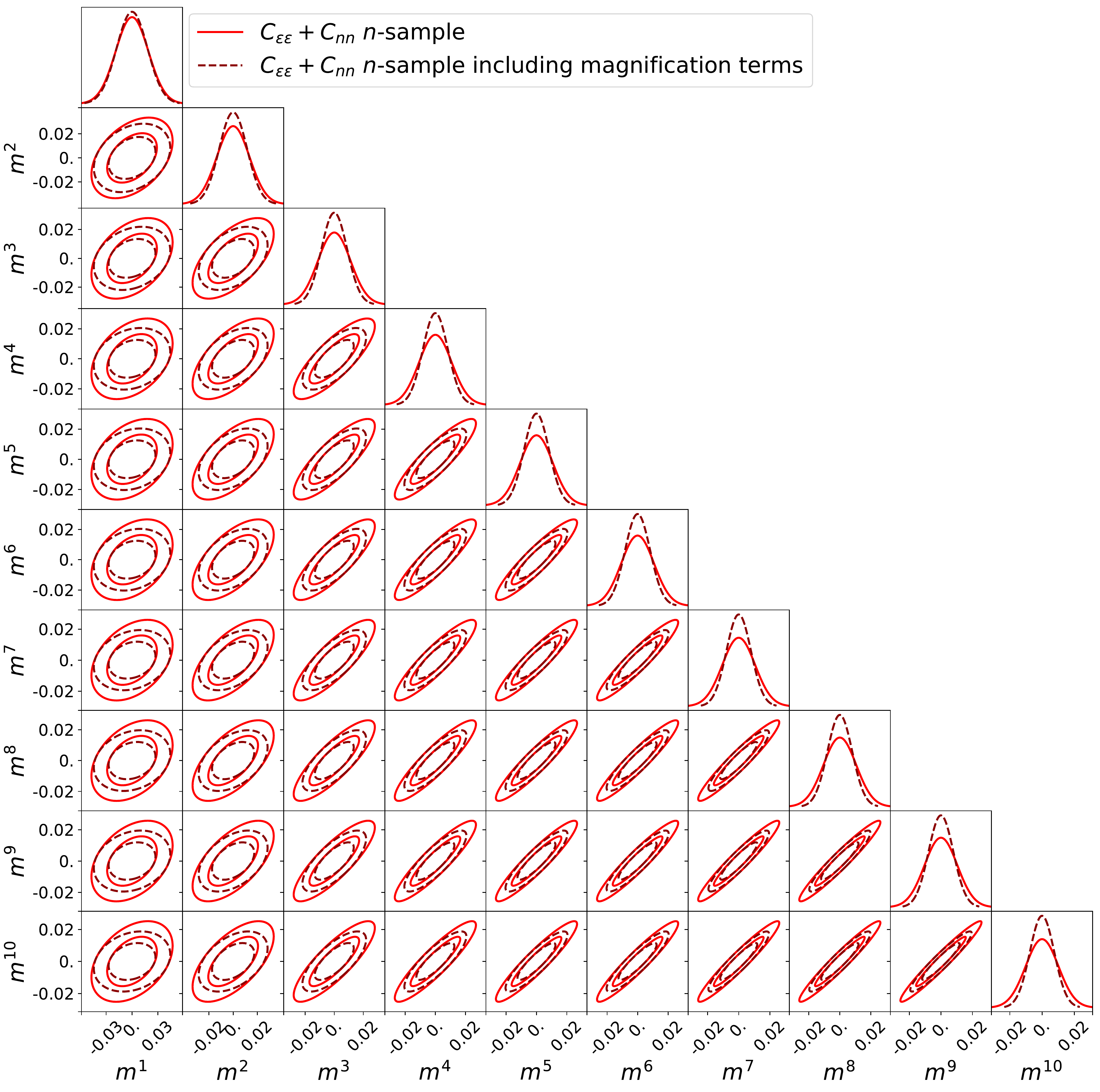}
  \caption{Constraints on the shear multiplicative bias parameters from a joint analysis of $C_{\rm{\epsilon\epsilon}}$ and $C_{\rm{nn}}$ (not including $C_{\rm{n\epsilon}}$) with and without magnification terms, where we have not applied the Gaussian prior detailed in Table \ref{tab:fiducial values}. $C_{\rm{\epsilon\epsilon}}$ is calculated for $\epsilon$-sample and $C_{\rm{nn}}$ is calculated for \textit{n}-sample. Including magnification only slightly improves the constraints, with a greater impact at higher redshift.}
  \label{fig:joint_shear_calibration}
\end{figure*}

Figure \ref{fig:joint_shear_calibration} shows the forecast constraints on the shear multiplicative bias parameters from our $C_{\rm{nn}}$ and $C_{\rm{\epsilon\epsilon}}$ analysis, with and without magnification terms, where $C_{\rm{\epsilon\epsilon}}$ is calculated for $\epsilon$-sample and $C_{\rm{nn}}$ for the \textit{n}-sample. Including magnification only slightly improves the constraints on the shear calibration parameters, with a greater effect at higher redshift. The 1$\sigma$ constraint on $m^1$ is improved by a factor of 1.06, $m^6$ by 1.3 and $m^{10}$ by 1.34 when including magnification. When $C_{\rm{nn}}$ is calculated using the $\epsilon$-sample the impact is similar, but less pronounced. These results show that including magnification is not particularly helpful for calibrating the shear measurement. However, the impact of magnification may be slightly improved when performing a full `3x2pt' analysis, where the clustering and shear are measured on the same patch of sky.

\subsection{Bias} \label{sec:bias results}

\begin{figure*}
  \centering
  \includegraphics[width=1.5\columnwidth]{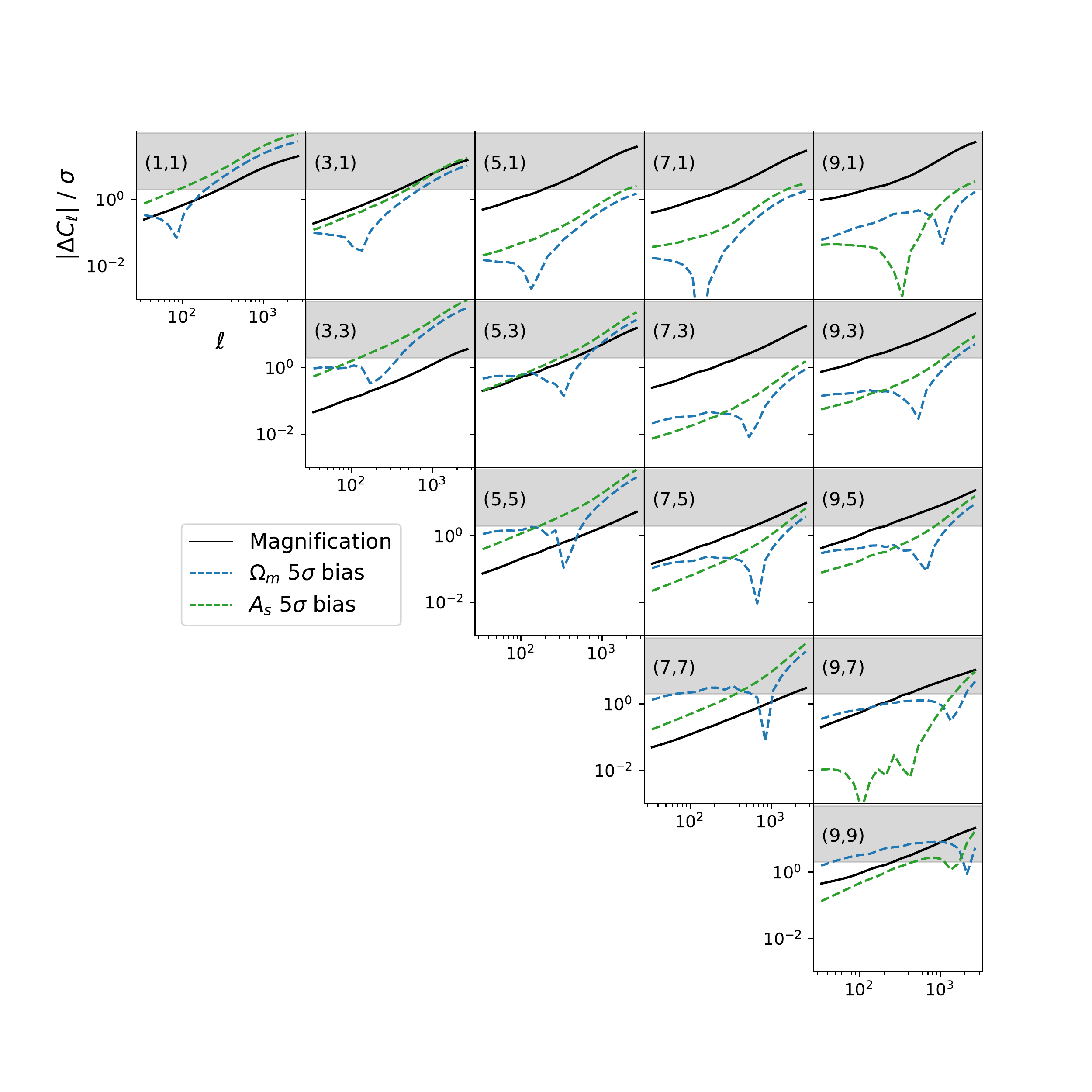}
  \caption{Absolute difference between $C_{\rm{nn}}$ with and without magnification, in terms of the uncertainty on $C_{\rm{nn}}$ without magnification, for \textit{n}-sample. The grey shaded region indicates where $C_{\rm{nn}}$ including magnification is more than $2\sigma$ away from $C_{\rm{nn}}$ without magnification. The dashed lines show the difference in $C_{\rm{nn}}$ when $\Omega_{\rm{m}}$ and $A_{\rm{s}}$ are altered by $5\sigma$.}
  \label{fig:bias_faint}
\end{figure*}

Recent works have shown that cosmological results from upcoming surveys such as LSST will be biased if the effects of weak lensing magnification are not included, due to improvements in statistical precision \citep{Duncan:2013haa, Cardona:2016qxn, Lorenz:2017iez, Thiele:2019fcu}. To examine this for our forecast, figure \ref{fig:bias_faint} shows the absolute difference between the clustering power spectra $C_{\rm{nn}}$ with and without magnification in terms of the $1\sigma$ uncertainty on the clustering power spectra without magnification. In this case the clustering power spectra have been calculated using \textit{n}-sample. The grey shaded region indicates where $C_{\rm{nn}}$ including magnification is more than $2\sigma$ away from $C_{\rm{nn}}$ without magnification. Particularly at high $\ell$ (small scales) $C_{\rm{nn}}$ including magnification significantly diverges from $C_{\rm{nn}}$ without magnification. It is worth noting that this result is influenced by the very small uncertainty on the clustering signal for a sample of such great depth. For a clustering signal with greater uncertainty the difference due to magnification in terms of the $1\sigma$ uncertainty would be reduced.

For qualitative comparison, we have also shown the impact of changing $\Omega_{\rm{m}}$ and $A_{\rm{s}}$ by $5\sigma$ in Fig. \ref{fig:bias_faint}. In all of the redshift bin combinations shown, the difference from including magnification is larger than or comparable to the difference from changing $\Omega_{\rm{m}}$ and $A_{\rm{s}}$ by $5\sigma$. This clearly indicates that not including magnification terms will catastrophically bias cosmological constraints from LSST. Additionally, the difference from not including magnification seems to mimic the behaviour of biasing $A_{\rm{s}}$ by $5\sigma$. This implies that not including magnification could particularly bias the constraints for $A_{\rm{s}}$, one of the parameters weak lensing is most sensitive to.

\begin{figure*}
  \centering
  \includegraphics[width=1.5\columnwidth]{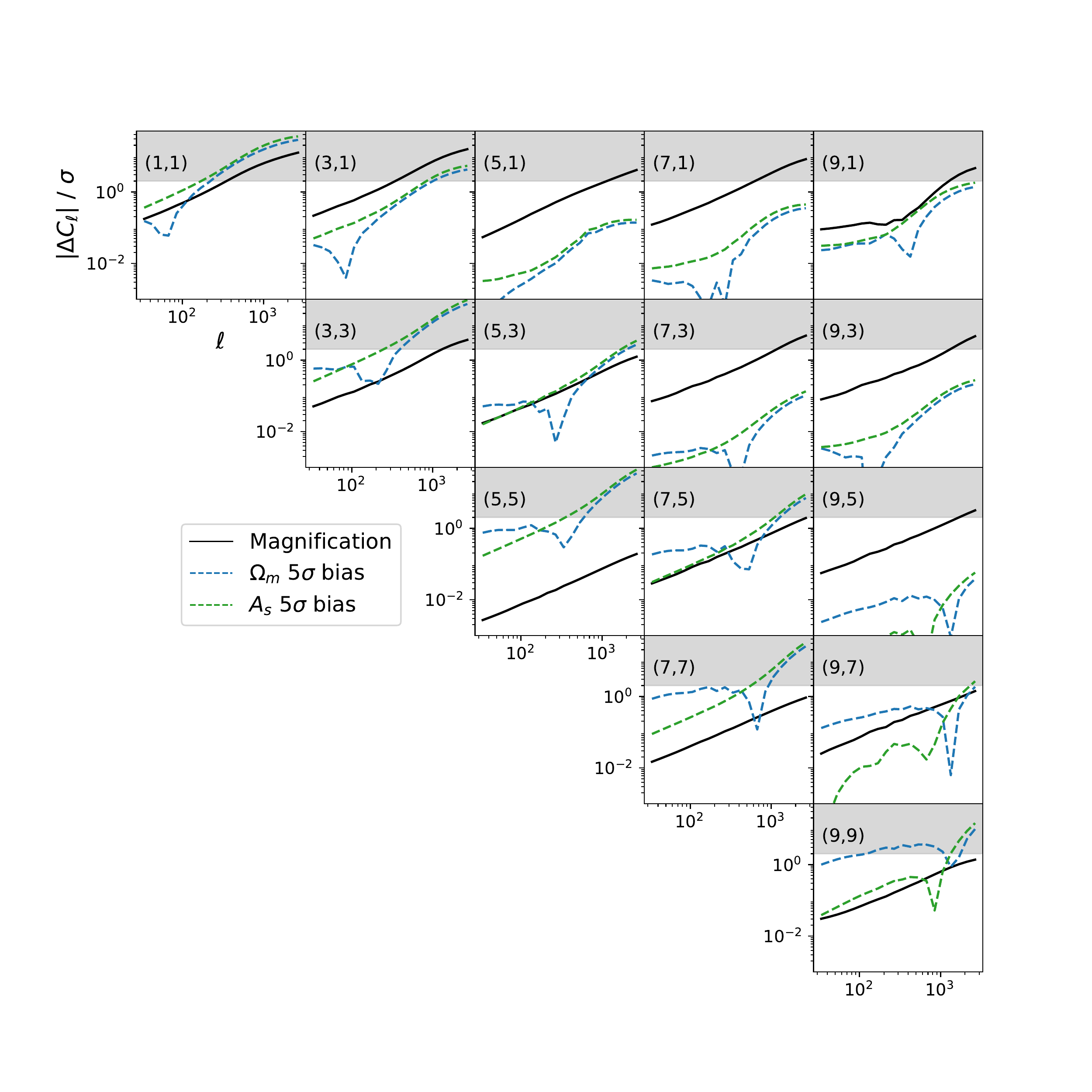}
  \caption{Absolute difference between $C_{\rm{nn}}$ with and without magnification, in terms of the uncertainty on $C_{\rm{nn}}$ without magnification, for $\epsilon$-sample. The grey shaded region indicates where $C_{\rm{nn}}$ including magnification is more than $2\sigma$ away from $C_{\rm{nn}}$ without magnification. The dashed lines show the difference in $C_{\rm{nn}}$ when $\Omega_{\rm{m}}$ and $A_{\rm{s}}$ are altered by $5\sigma$.}
  \label{fig:bias_gold}
\end{figure*}

Figure \ref{fig:bias_gold} shows the absolute difference between the clustering power spectra $C_{\rm{nn}}$ with and without magnification in terms of the $1\sigma$ uncertainty on the clustering power spectra without magnification, where the clustering power spectra have been calculated using the $\epsilon$-sample. In this case the difference from including magnification is not as large as for \textit{n}-sample, however in most redshift bin combinations is still comparable or larger than the differences from changing $\Omega_{\rm{m}}$ and $A_{\rm{s}}$ by $5\sigma$. 

\section{Conclusions} \label{sec:Conclusions}

Previous works have shown that upcoming results from surveys such as LSST and \textit{Euclid} will be biased if the effects of weak lensing magnification are not included \citep{Duncan:2013haa, Cardona:2016qxn, Lorenz:2017iez, Thiele:2019fcu}. In this work we forecast whether including weak lensing magnification as a complementary probe can additionally improve the precision of the LSST galaxy clustering constraints. We determined this using the Fisher matrix formalism, where our theory datavector included galaxy clustering and the galaxy luminosity function. To calculate the galaxy clustering and the galaxy luminosity function, we employed a halo model, detailed in \cite{Cacciato2013MNRAS.430..767C}. We defined two mock LSST galaxy samples from the LSST DC2 simulations \citep{Korytov:2019xus} for use in our forecast; a sample which corresponds to the LSST gold sample where the $i$ band magnitude is less than 25.3 (intended to be used for the weak lensing shear measurement), and a very deep sample where the $i$ band magnitude is less than 26.5.

We found that weak lensing magnification provides little additional information as a complementary probe for LSST. For a galaxy clustering analysis using the LSST gold sample we found that including magnification increased the Figure of Merit (FoM) for the set of cosmological parameters $\Omega_{\rm{m}}$, $h_0$, $\Omega_{\rm{b}}$, $n_{\rm{s}}$, $A_{\rm{s}}/10^{-9}$, $w$ and $w_{\rm{a}}$ by a factor of 1.08. When using the deep galaxy sample we found that magnification increased the FoM by a factor of 1.45. In terms of the precision of the $\Omega_m$ constraints, we found for a galaxy clustering analysis using the LSST gold sample that including magnification increased the 1$\sigma$ precision by a factor of 1.03, using the deep sample we found a factor increase of 1.3. These results show that including magnification is more beneficial for deeper samples.

The effect of including magnification would be smaller in a combined galaxy clustering and cosmic shear analysis because magnification provides similar information to that of cosmic shear. However, we investigated the impact of including magnification on the calibration of the shear measurement. We found that including magnification only slightly improves the constraints on the shear calibration parameters.

While this forecast is more realistic than many to date, as it includes LSST mock catalog data and a flexible galaxy bias model, it still relies on a number of simplified assumptions about magnification. Firstly, the magnification modelling assumes that the galaxy sample is purely flux limited. Often galaxies are also selected based on their signal-to-noise ratio, colours and morphology which complicates the magnification modelling \citep{hildebrandt10.1093/mnras/stv2575}. Secondly, there are a large number of systematics associated with the magnification measurement such as dust attenuation, variable survey depth, star-galaxy separation and the blending of galaxy images \citep{Hildebrandt2013MNRAS.429.3230H,Morrison:2015aia,Thiele:2019fcu}. We included a multiplicative factor in our modelling of the clustering power spectra in order to incorporate these effects, but more detailed modelling is likely required. For example, we could have marginalised over the faint end slopes of the number counts $\alpha^i$, which are required to compute the magnification power spectra. We chose to fix them, since at least for the gold sample it should be comparatively easy to explore the luminosity function beyond the magnitude limit, so  measurement errors on $\alpha^i$ can be expected to be very small. This forecast could therefore be considered a best case scenario for magnification, and even in this scenario we found that including magnification has little impact. However, we also confirmed that not including magnification will strongly bias cosmological results from LSST, so must be modelled.

\section*{Acknowledgements}
This paper has undergone internal review in the LSST Dark Energy Science Collaboration. We would like to thank the internal reviewers David Alonso, Sukhdeep Singh and Marina Ricci for their insightful comments. We also thank Hendrik Hildebrandt for comments on the manuscript, and Christopher Duncan and Harry Johnston for useful discussions. CM was supported by the Spreadbury Fund, Perren Fund, IMPACT Fund, and by the European Research Council under grant 770935. CM acknowledges travel support provided by STFC for UK participation in LSST through grant ST/N002512/1. MCF and HH acknowledge support from the Netherlands Organisation for Scientific Research (NWO) through grant 639.043.512. BJ acknowledges support by the UCL Cosmoparticle Initiative. AK acknowledges support through STFC grant ST/S000666/1. SJS acknowledges support from DOE grant DESC0009999 and NSF/AURA grant N56981C. We acknowledge the use of \textsc{CosmoSIS}\footnote{https://bitbucket.org/joezuntz/cosmosis/wiki/Home} and \textsc{hmf}\footnote{https://github.com/halomod/hmf}, and thank their authors for making these products public. 

The DESC acknowledges ongoing support from the Institut National de 
Physique Nucl\'eaire et de Physique des Particules in France; the 
Science \& Technology Facilities Council in the United Kingdom; and the
Department of Energy, the National Science Foundation, and the LSST 
Corporation in the United States.  DESC uses resources of the IN2P3 
Computing Center (CC-IN2P3--Lyon/Villeurbanne - France) funded by the 
Centre National de la Recherche Scientifique; the National Energy 
Research Scientific Computing Center, a DOE Office of Science User 
Facility supported by the Office of Science of the U.S.\ Department of
Energy under Contract No.\ DE-AC02-05CH11231; STFC DiRAC HPC Facilities, 
funded by UK BIS National E-infrastructure capital grants; and the UK 
particle physics grid, supported by the GridPP Collaboration.  This 
work was performed in part under DOE Contract DE-AC02-76SF00515.

\section*{Data Availability}

The cosmoDC2 catalog is publicly available here: \url{https://portal.nersc.gov/project/lsst/cosmoDC2/_README.html}




\bibliographystyle{mnras}
\bibliography{magnification} 




\appendix

\section{Volume Complete Cut for Galaxy Luminosity Function Covariance} \label{sec:vol_complete_cut}

\begin{figure*}
  \centering
  \includegraphics[width=1.5\columnwidth]{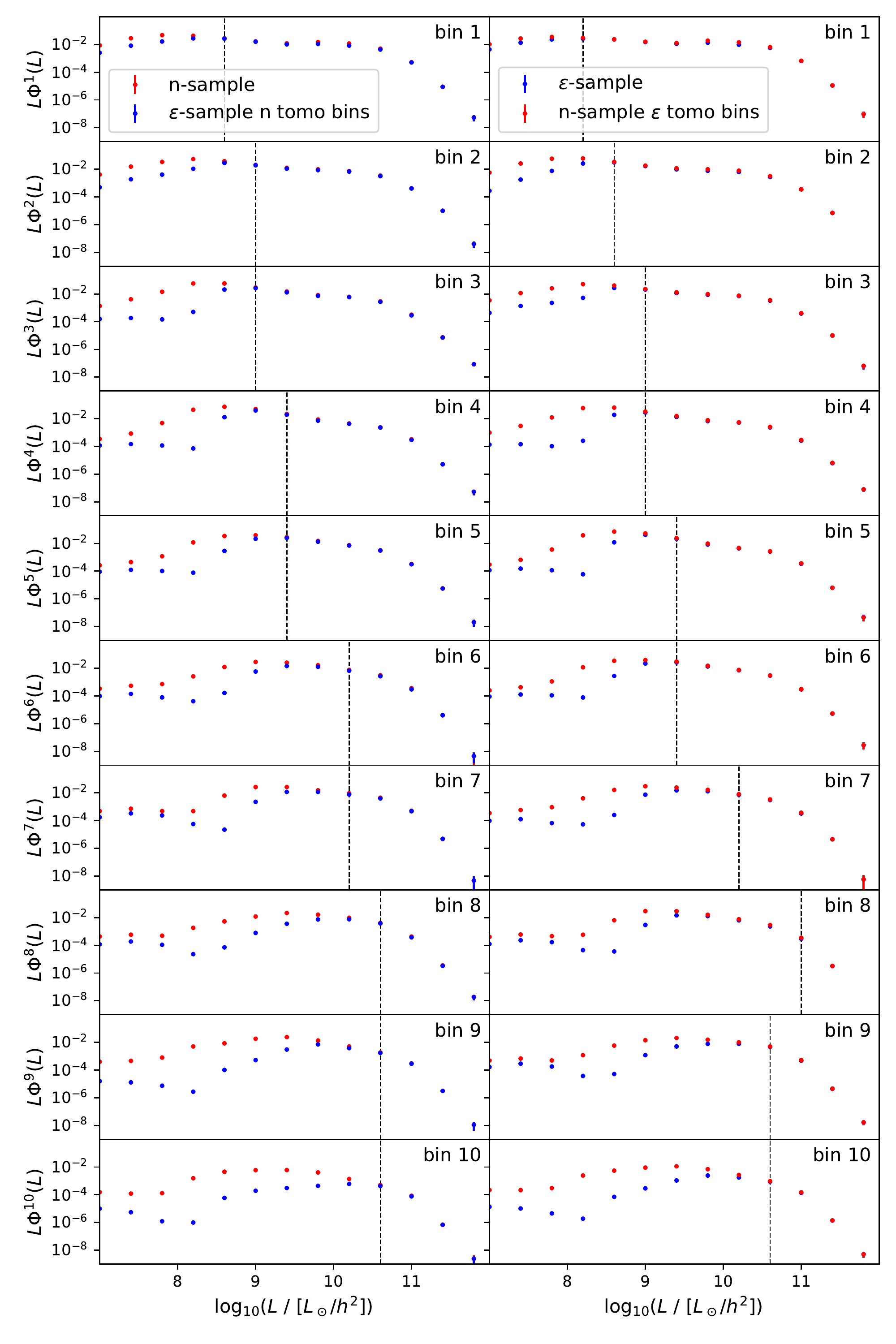}
  \caption{Measured luminosity functions $\Phi^i(L)$ for each photometric redshift bin in \textit{n}-sample (left) and $\epsilon$-sample (right), with associated bootstap errors. The dotted lines show the location of the luminosity cuts to \textit{n}-sample (left) and $\epsilon$-sample (right) to make sure they are volume complete, and do not introduce a bias. $\epsilon$-sample n tomo bins refers to the $\epsilon$-sample being binned into the \textit{n}-sample redshift bins, and \textit{n}-sample $\epsilon$ tomo bins refers to the \textit{n}-sample being binned into $\epsilon$-sample redshift bins. $L\Phi^i(L)$ has units of $h^3/\mathrm{Mpc}^3$.}
  \label{fig:volume_complete_cuts}
\end{figure*}

A deeper galaxy sample will be volume complete to lower luminosities, so when the luminosity function of a shallower sample diverges from the luminosity function of a deeper sample, we know the shallower sample has ceased to be volume complete. We can therefore determine the volume complete luminosity cut for the $\epsilon$-sample by finding where it diverges from the \textit{n}-sample. Our divergence condition is
\begin{equation}
  \frac{|\Phi^i_\epsilon(L)-\Phi^i_{n(\epsilon)}(L)|}{\Phi^i_\epsilon(L)}>0.2 \ ,
\end{equation}
where $\Phi^i_\epsilon$ is the luminosity function for the $\epsilon$-sample and $\Phi^i_{n(\epsilon)}$ is the luminosity function for the \textit{n}-sample, where the \textit{n}-sample has been binned using the $\epsilon$-sample tomographic bins. We cut $\Phi^i_\epsilon$ when there is a difference of 20\% from the deeper sample $\Phi^i_{n(\epsilon)}$. This value was found to cut $\Phi^i_\epsilon$ before it significantly diverged from the deeper sample whilst allowing for small deviations, see the right panel of Fig. \ref{fig:volume_complete_cuts}.

Since we did not have a sample deeper than the \textit{n}-sample available to us, we made a more stringent volume complete cut on the \textit{n}-sample luminosity function based on where the luminosity function of our shallower sample $\epsilon$-sample diverged. If the shallower sample is volume complete we can be sure that the deeper sample is also volume complete. In this case our divergence condition is
\begin{equation}
  \frac{|\Phi^i_n(L)-\Phi^i_{\epsilon(n)}(L)|}{\Phi^i_n}>0.2 \ ,
\end{equation}
where $\Phi^i_n$ is the luminosity function for \textit{n}-sample and $\Phi^i_{\epsilon(n)}$ is the luminosity function for $\epsilon$-sample, where $\epsilon$-sample has been binned using the \textit{n}-sample tomographic bins. While this luminosity cut enforces that \textit{n}-sample is volume complete, using a shallower sample means that the cut is much more conservative than necessary.

\section{Fisher Matrix Stability} \label{sec: Fisher matrix stability}
High-dimensional Fisher matrices can be unstable. Here we detail the steps taken to ensure the stability of our Fisher matrices and hence the robustness of our results.

The derivatives in eq. (\ref{eq:fisher_sum}) are calculated numerically using a method of numerical differentiation called a 5-pt stencil. This method requires the pipeline to be evaluated at 4 points around the model parameter's fiducial value (5 points including the fiducial value). The separation between these points is referred to as the step size. If the step size is too large the Fisher matrix fails to capture the curvature of the likelihood function about the peak and if it is too small numerical difficulties can arise. Therefore when using Fisher matrices it is vital to verify whether the step size is appropriate, otherwise any results are meaningless.

We verify our step sizes in 1 dimension by fixing all but one model parameter. We then calculate the 1D likelihood using a Fisher matrix with a specified step size and by sampling the likelihood function directly. If the 1D likelihoods match we know we are using a reasonable step size when calculating our Fisher matrix. We sample the likelihood function directly using a simulated datavector generated at the Fisher matrix fiducial values and a grid sampler. Grid samplers evaluate the likelihood at a specified set of grid points. Since we are assuming a Gaussian Likelihood when calculating our Fisher matrix (eq. (\ref{eq:fisher_sum})) we are only interested in whether the standard deviation $\sigma$ of the likelihood calculated using the Fisher matrix matches the $\sigma$ of the likelihood from sampling directly using a grid sampler. 

\begin{figure*}
  \rotatebox{90}{
  \begin{minipage}{0.9\textheight}
  \includegraphics[width=1.0\linewidth]{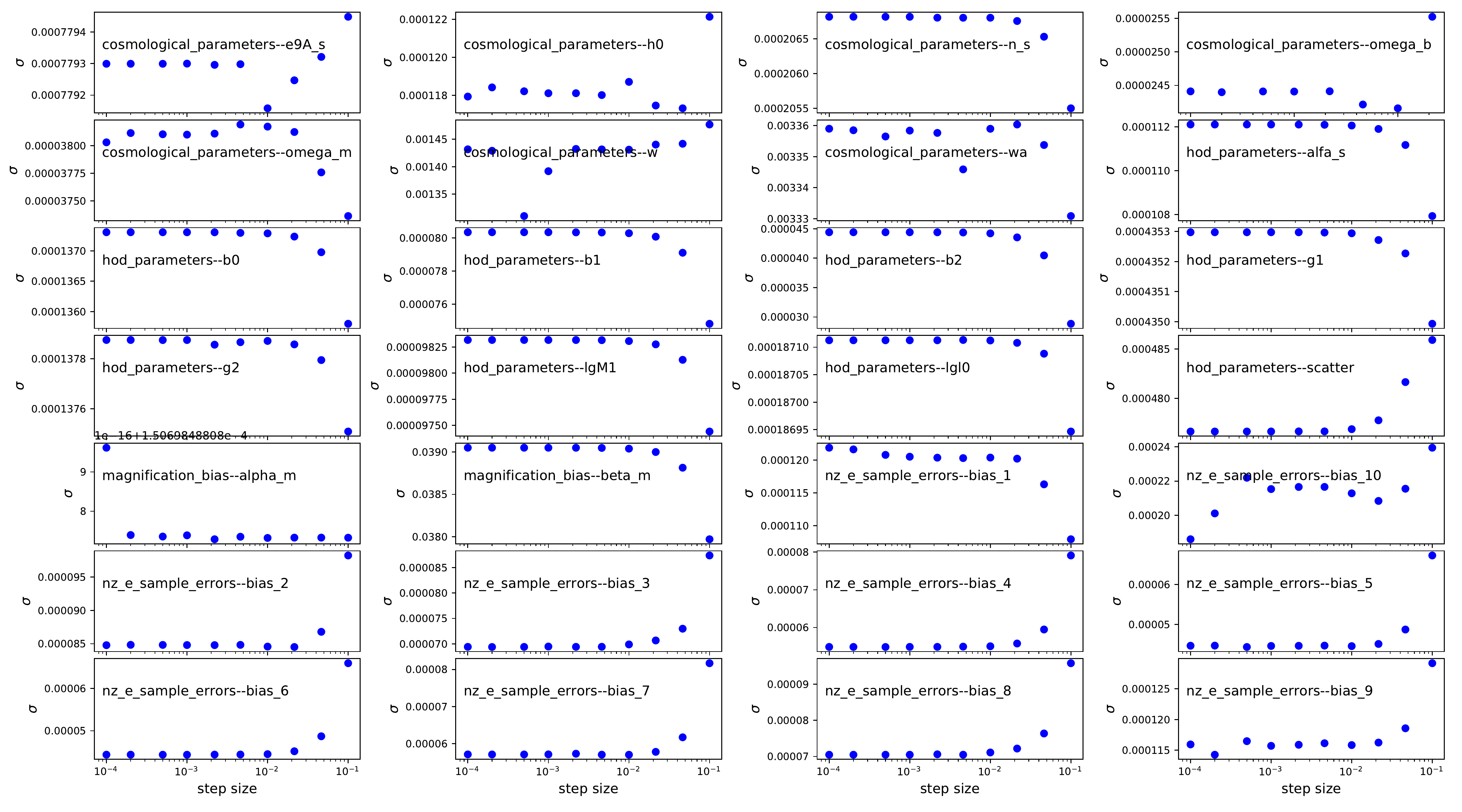}
  \caption{Standard deviation of the 1D Fisher likelihoods against the step size used to calculate the derivatives in the Fisher matrix. The chosen step size should be in the plateau region where the Fisher matrix is actually capturing the shape of the likelihood.}
  \label{fig:fisher variance}
  \end{minipage}}
  \end{figure*}

\begin{figure*}
    \rotatebox{90}{
    \begin{minipage}{0.9\textheight}
    \includegraphics[width=1.0\linewidth]{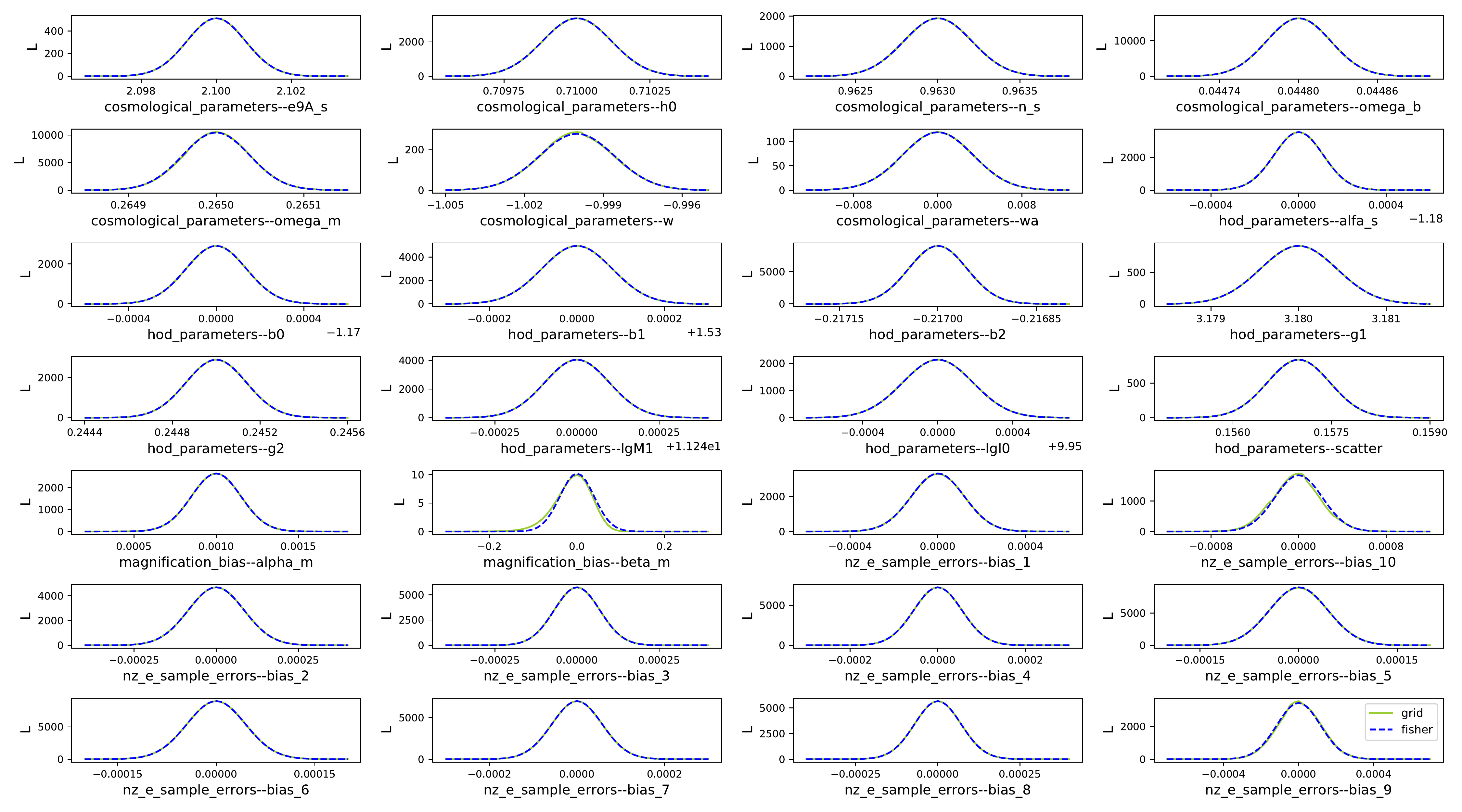}
    \caption{Comparison of the 1D likelihoods from the Fisher matrix calculated using the selected step size against the 1D likelihoods from sampling the likelihood directly using a grid sampler.}
    \label{fig:fisher matched}
    \end{minipage}}
    \end{figure*}

Figure \ref{fig:fisher variance} shows the $\sigma$ of the 1D likelihood calculated using the Fisher matrix for different choices of step size. These plots show that as the step size decreases the $\sigma$ of the 1D likelihood reaches a plateau, where the step size is actually capturing the shape of the likelihood, before becoming unstable (see subplot for the photometric redshift bias parameter for redshift bin 10). We therefore select a step size in the range where the $\sigma$ of the Fisher likelihood is stable. Figure \ref{fig:fisher matched} shows the Fisher likelihoods generated using the selected step sizes overlaid with the likelihood from the grid sampler to verify that they match. For the case of the magnification bias parameter $b_{\rm{m}}$ the Fisher and grid likelihoods do not match. This is because when calculating the Fisher matrix we assume that the likelihood is Gaussian, and the likelihood of $b_{\rm{m}}$ from direct sampling is clearly not Gaussian. This is a limitation of the Fisher matrix approach.

\begin{figure*}
    \centering
    \includegraphics[width=1.5\columnwidth]{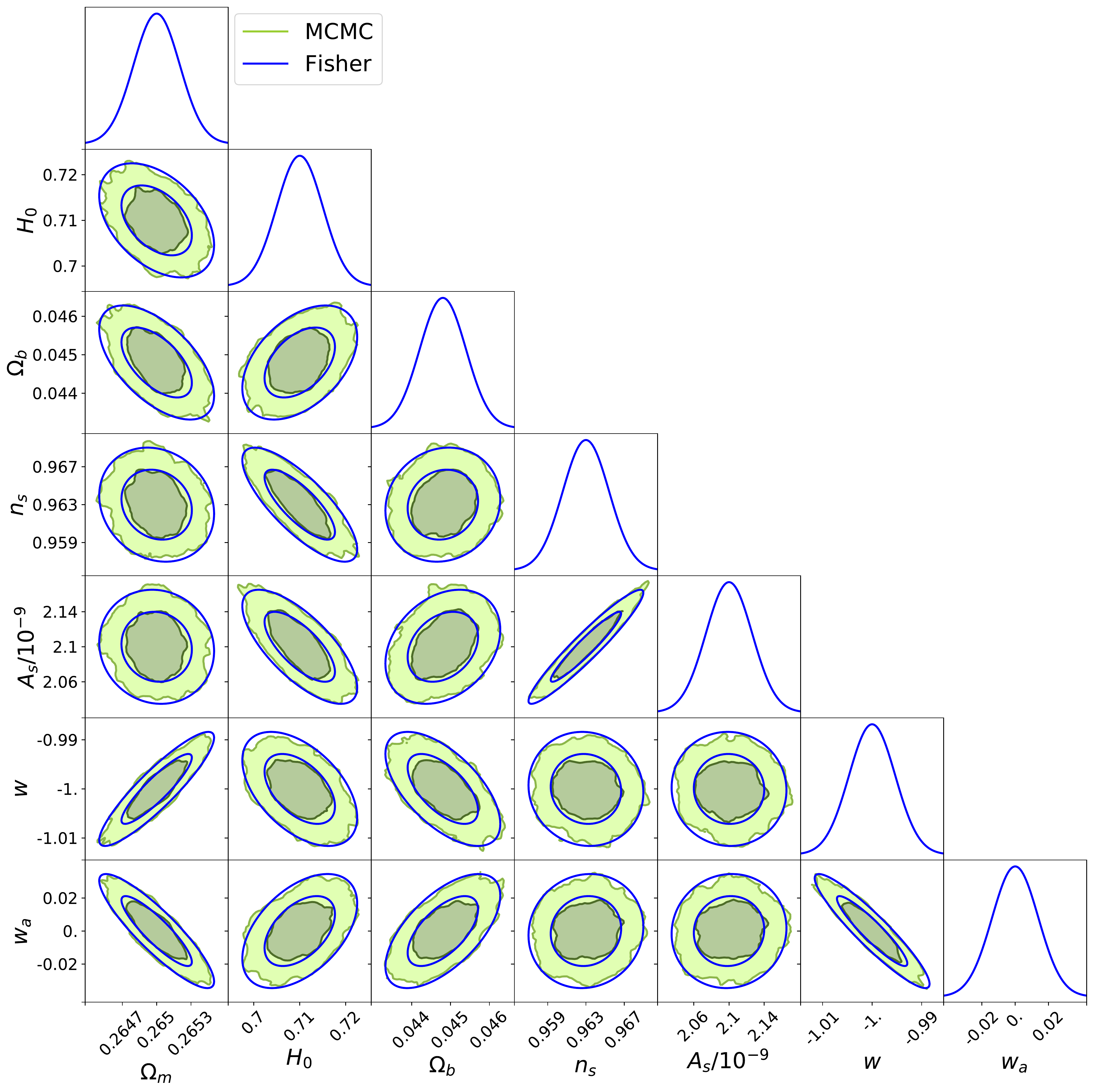}
    \caption{Comparison of the constraints on the cosmological parameters used in this analysis when found using and MCMC or a Fisher matrix. All other parameters have been fixed.}
    \label{fig:MCMC_comparison}
  \end{figure*}

We additionally check the Fisher step sizes for the cosmological parameters, by varying all the cosmological parameters at once and exploring the multivariate posterior with Markov Chain Monte Carlo (MCMC) sampling\footnote{the MCMC we use is emcee \citep{2013PASP..125..306F}}. Figure \ref{fig:MCMC_comparison} shows a comparison between the constraints obtained from the MCMC and the Fisher matrix. They match well and show that our Fisher matrix is adequately capturing the shape of the likelihood.

Figures \ref{fig:fisher variance} and \ref{fig:fisher matched} show only an example case for the parameters used to generate the $C_{\mathrm{nn}}$ Fisher matrix for $\epsilon$-sample. However, the step sizes have been verified using this method for every Fisher matrix referred to in the results section.

\bsp	
\label{lastpage}
\end{document}